\documentclass[twocolumn]{article2}%

\usepackage{newtxtext,newtxmath}
\usepackage{natbib}
\usepackage[justification=justified]{caption}
\usepackage{bm}
\usepackage{booktabs}
\usepackage{xr}

\renewcommand{\d}{\,\mathrm{d}}
\DeclareMathOperator{\e}{\mathrm{e}}
\DeclareMathOperator{\E}{\mathbb{E}}
\DeclareMathOperator{\V}{\mathbb{V}}
\newcommand{\U}{\mathbb{U}}
\newcommand{\N}{\mathbb{N}}

\renewcommand{\P}{\mathbb{P}}
\newcommand{\Pj}{P_{\mathrm{jump}}}

\newcommand{\red}[1]{{\color{red}{#1}}}

\begin{document}

\title[MCMC mixing efficiency]{Mixing efficiency of trans-model Markov chain Monte Carlo
algorithms with applications in Bayesian phylogenetics}%

\author*[1]{\fnm{Xiyun} \sur{Jiao} (orcid: 0009-0006-3924-985X)} \email{jiaoxy@sustech.edu.cn}%

\author[2]{\fnm{Thomas} \sur{Flouris} (orcid: 0000-0002-8474-9507)}\email{t.flouris@ucl.ac.uk}
\author*[2]{\fnm{Ziheng} \sur{Yang} (orcid: 0000-0003-3351-7981)} 
\email{z.yang@ucl.ac.uk}

\affil*[1]{Department of Statistics and Data Science, China Southern University of
   Science and Technology, Shenzhen, Guangdong 518055, China} %

\affil*[2]{Department of Genetics, Evolution, and Environment, University College
   London, Gower Street, London WC1E 6BT, UK} %

\abstract{Trans-model Markov chain Monte Carlo (MCMC) algorithms are widely used in
Bayesian inference, and are particularly important in Bayesian phylogenetics where
phylogenetic trees represent different statistical models.  While the algorithm allows
great flexibility, its mixing efficiency can vary hugely, and is poorly understood. 
Here we use mathematical analysis and simulation to explore the mixing efficiency of
trans-model MCMC proposals, including the model-proposal probabilities and the proposal
kernel for model parameters. Our analysis confirms the intuition that one should
preferentially propose models with high posterior probabilities, and propose parameter
values from the posterior as much as possible.  Our results provide guidelines for
constructing efficient trans-model MCMC algorithms.  The principles are applied to MCMC
algorithms in phylogenetic reconstruction using two real datasets for primates and
mammals. }%

\maketitle

\section{Introduction}

Since its introduction into molecular phylogenetics in the 1990s \citep{Rannala1996,
Yang1997, Mau1997, Li2000}, Bayesian inference has become one of the most popular
statistical methods in the field \citep{Chen2014, Yang2014}.  Many sophisticated models
of molecular sequence evolution have been implemented in popular programs such as
\textsc{MrBayes} \citep{Ronquist2012}, \textsc{beast} \citep{Bouckaert2014}, and
\textsc{PhyloBayes} \citep{Lartillot2009}, used to estimate parameters and comparing
models of sequence evolution, and to infer species phylogenies. Phylogeny reconstruction
is an example of Bayesian model selection, as the phylogenetic trees (together with the
model of sequence evolution) specify the likelihood function and correspond to different
likelihood models, while branch lengths in each tree (as well as parameters
characterizing the evolutionary model) are parameters in the model. Computation is
achieved by trans-model Markov chain Monte Carlo (MCMC) algorithms \citep{Metropolis1953,
Hastings1970, Green1995}, which move between trees, and the frequency at which the MCMC
visits each tree is an estimate of its posterior probability.

It has been noted that posterior probabilities for inferred trees or clades are often
$\sim$100\% in modern phylogenomic datasets \cite[e.g.,][]{Yang2018}, prompting some
researchers to question the utility of posterior probabilities for trees when the
dataset is large and the model is misspecified \cite[e.g.,][]{Thomson2022}.  These are
important questions, as are the choice of the model of sequence evolution and
specification and impact of the prior \citep{Yang2014}, but these are beyond the scope
of this paper.  Here we focus on the efficiency of MCMC sampling from the posterior when
the data, prior and model are fixed.

A number of constructions have been introduced for cross-model MCMC moves.  In the
\textit{product-space} construction \citep{Carlin1995}, the state of the Markov chain
includes all parameters across all models.  When the chain is in one model, parameters
of other models have no influence on the likelihood or posterior, but are treated as
pseudo-parameters and updated as well, using pseudo-priors.  Green \citep{Green1995}
developed the idea of dimension-matching to allow moves between models of different
sizes in the reversible-jump MCMC (rjMCMC) algorithm.  The \textit{composite-space}
framework \citep{Godsill2001} allows parameters to overlap arbitrarily between models,
and includes rjMCMC and the product-space construction as special cases.  The
\textit{saturated-space} approach \citep{Brooks2003} is similar and augments the state
space of the `small' models to achieve the same dimension as the `largest' model.
Different trans-model sampling schemes were compared in \citep{Dellaportas2002}.  Several
other algorithms for cross-model inference \citep{Grenander1994,Stephens2000,Cappe2003}
may be considered particular versions of rjMCMC.  For within-model inference problems, a
number of authors discussed the benefits of using local information of the posterior to
guide the proposals \citep{Zanella2020}.  For example, the local gradient can be used to
guide the proposal toward high-probability regions, as in the Metropolis-adjusted
Langevin algorithm (MALA, \citep{Roberts1998}) and Hamiltonian Monte Carlo (HMC)
\citep{Neal2011, Girolami2011}.  More recently, HMC is being adapted to propose moves
between models \citep{Nishimura2020}.


In Bayesian phylogenetics, branch-swapping algorithms developed in parsimony and
likelihood tree search, such as nearest-neighbor-interchange (NNI) and subtree-pruning and
regrafting (SPR) \citep{Swofford1996}, have been adapted into MCMC proposal algorithms to
introduce local stochastic alterations to the current tree \citep{Lakner2008, Hohna2008,
   Yang2014}.  There have been efforts to design locally-informed MCMC moves that better
reflect the posterior than blind proposals that select candidate trees with uniform
probabilities.  For example, one can select preferentially short internal branches for
tree perturbation \citep{Rannala2017}, and use weights to sample target branches for
regrafting, based on the parsimony score of the resulting tree \citep{Yang2014, Zhang2020},
conditional probabilities of clade \citep{Hohna2012}, or posterior probabilities for
splits (i.e., bi-partitions of species defined by internal branches on the tree) estimated
during a pilot run \citep{Meyer2021}.

While the efficiency of within-model MCMC \cite[i.e., the Metropolis-Hastings (MH)
algorithm,][]{Metropolis1953, Hastings1970} has received much attention in the literature,
especially concerning the optimal step size of Gaussian sliding-window moves
\citep{Gelman1996} or the choice of proposal kernels \citep{Yang2013, Thawornwattana2018BA},
there is a lack of results concerning efficient MCMC algorithms that move between models
\citep{Zanella2020}.  Here efficiency of cross-model MCMC may be defined as the ratio of
the variance of the estimate of a posterior model probability based on an independent
sample to the variance from an MCMC sample of the same size \citep{Peskun1973, Gelman1996}.
While trans-model MCMC algorithms allow great flexibility \citep{Green1995}, designing
trans-model moves with good mixing properties is perhaps more art than science
\citep{Felsenstein2004}.  Many questions remain concerning the efficiency of trans-model
MCMC algorithms in general, and cross-tree moves in Bayesian phylogenetics in particular.
For example, does the most efficient cross-model chain always have the highest model-jump
probability or acceptance rate ($\Pj$), and does the highest $\Pj$ imply maximum
efficiency?  Is it good to propose parameters of the new model from the prior? If two
models are nested, is it good to propose parameters around the point of model overlap
where the likelihood is identical between the two models?  Does the inclusion of
within-model moves updating the parameters always improve the efficiency of the
cross-model algorithm?  How should one divide the computational effort between within- and
cross-model moves?

We address those questions in this paper.  We analyze two simple examples of trans-model
MCMC, with uniform and normal target distributions, respectively, before developing a
theory about the general case of comparing two models with one of them having parameters.
We then study several variants of the SPR algorithm for stochastic tree search, to
illustrate the theory developed through the simple examples.  We focus on two features of
a trans-model MCMC algorithm: (i) the model-jump probability ($\Pj$) and (ii) the
efficiency of the MCMC sample for estimating the posterior probability of a model.

\section{Overview of trans-model MCMC algorithms}

\subsection{Trans-model MCMC algorithm}

Suppose there are $K$ candidate models $\{H_k\}, k=1, \dots, K$.  Model $H_k$ has a
vector of $n_k$ unknown parameters $\theta_k \in \Theta_k$.  Let $X$ be the data, and
$p(X | k, \theta_k)$ be the likelihood under model $k$.  We assign the prior $p(k,
\theta_k) = p(k) p(\theta_k|k)$.  The joint posterior of the model and parameters is
then
\begin{equation}  \label{eq:posterior-joint}
 p(k, \theta_k|X) \propto p(k) p(\theta_k|k) p(X|k, \theta_k).
\end{equation}
We have $p(k, \theta_k|X) = p(k|X) p(\theta_k|X,k)$.  We consider MCMC algorithms that
sample from the joint posterior of eq.~\ref{eq:posterior-joint}.  Let $\omega = (k,
\theta_k)$ be the state of the Markov chain, with the state space $\Omega = \cup_{k=1}^K
(\{k\}, \Theta_k)$, which is the union of the subspaces for all $K$ models.  We write
$p(k, \theta_k|X) \equiv \pi(k, \theta_k)$, $p(k|X) \equiv \pi(k)$, and $p(\theta_k|k,X)
\equiv \pi(\theta_k|k)$, so that $\pi(\omega) \equiv \pi(k, \theta_k) = \pi(k)
\pi(\theta_k|k)$.

To propose a move from $\omega = (k, \theta_k)$ to $\omega' = (k', \theta_{k'} )$, we
propose a new model $k'$ with probability $q_{kk'}$, and parameters for the new model
from a proposal kernel $q(\theta_{k'}| k, \theta_k, k')$.  The move is accepted with
probability $\alpha(\omega, \omega') = \mathrm{min}\{1, A(\omega, \omega') \}$, where
\begin{equation} \label{eq:rjmcmc-accept}
   \begin{aligned}
      \small
      A(\omega, \omega')
      &= \frac{q(k,\theta_{k}| k', \theta_{k'})}{q(k',\theta_{k'}| k,
         \theta_k)}
      \times
      \frac{\pi(k', \theta_{k'})}{\pi(k, \theta_k)} \\
      &= \frac{q_{k' k}}{q_{k k'}}
      \cdot
      \frac{q(\theta_{k}| k', \theta_{k'}, k)}{q(\theta_{k'}| k, \theta_k, k')}
      \times
      \frac{\pi(k', \theta_{k'})}{\pi(k, \theta_k)}.      
   \end{aligned} 
\end{equation}
The Hastings ratio, $\frac{q(k,\theta_{k}| k', \theta_{k'})}{q(k',\theta_{k'}| k,
   \theta_k)}$, factors into two components: the model proposal probability ratio,
$\frac{q_{k' k}}{q_{kk'}}$, and the parameter proposal ratio, $\frac{q(\theta_{k}| k',
   \theta_{k'}, k)}{q(\theta_{k'}| k, \theta_k, k')}$.  Here models $k$ and $k'$ may have
different sizes ($n_k \ne n_{k'}$).

The resulting Markov chain, $\{ \omega^{(t)} \}$, has the transition kernel
{\scriptsize
\begin{multline} \label{eq:transition-kernel}
   p(k', \theta_{k'} | k, \theta_k) \\
   = %
   \begin{cases}
      \setlength\arraycolsep{10em}
      q(k',\theta_{k'} | k,\theta_k) \alpha(k',\theta_{k'} | k,\theta_k), & \text{if } \omega' \ne \omega, \\
      1 - \sum_{k'} \int q(k',\theta_{k'} | k,\theta_k) \alpha(k',\theta_{k'} | k,\theta_k) \d\theta_{k'}, & \text{if } \omega' = \omega,
   \end{cases}
\end{multline}
}
The point mass at $\omega' = \omega$ is due to rejection of proposals.

\subsection{The model-jump probability}

We define the \textit{model jump probability} as an average over the target distribution
of the model and parameters:
\begin{equation} \label{eq:pjump}
   \Pj = \sum_{k=1}^K \pi_k (1-p_{kk}^*),
\end{equation}
where
\begin{equation} \label{eq:pkk}
   \small%
   p_{kk'}^* = \int_{\Theta_k} \pi(\theta_k|k)
   \int_{\Theta_{k'}} p(k', \theta_{k'} | k, \theta_k) \, \d \theta_{k'} \,\d \theta_k .
\end{equation}
Note that $0\le p_{kk'}^* \le 1$, $\sum_{k'=1}^K p_{k k'}^* = 1$, and $\sum_{k=1}^K \pi_k
p_{k k'}^* = \pi_{k'}$.  Here $p_{kk'}^*$ may be interpreted as the average probability of
moving into model $k'$ in the next step given that the chain is currently in model $k$,
even though the sequence of model indices, $\{ k^{(t)} \}$, may not constitute a Markov
chain.

In the special case where there are no parameters in any of the models, we have
\begin{equation}
   \Pj = \sum_{k=1}^K \pi_k (1-p_{kk}),
\end{equation}
where $p_{k k'}$ is the transition probability from $k$ to $k'$.

\begin{theorem} 
$\Pj$ of eq.~\ref{eq:pjump} has an achievable upper bound
\begin{equation} \label{eq:pjump-bound}
      \Pj \le 2 (1 - \max\{\pi_k\}),
\end{equation}
where $\max\{\pi_k\}$ is the posterior probability of the most probable model (that is,
the maximum \textit{a posteriori} or MAP model).
\end{theorem}
A proof is provided in SI text 1.

\subsection{The mixing efficiency of trans-model MCMC}

Given an MCMC sample $\{ \omega^{(t)} \}$, $t=1,\cdots,N$, the expectation of any
integrable function $f(\omega)$ with respect to $\pi$,
\begin{equation} \label{eq:I}
I = \E_\pi(f) = \int_\Theta f(\omega) \pi(\omega) \d \omega,
\end{equation}
can be estimated by the sample mean
\begin{equation} \label{eq:Iapprox}
  \tilde I = \frac{1}{N} \sum_{t=1}^N f(\omega^{(t)}).
\end{equation}
This has the asymptotic variance
\begin{equation} \label{eq:nu}
  \nu = \lim_{N\to \infty} N \V\{\tilde I \} = \nu_f [1 + 2(\rho_1 + \rho_2 + \cdots)],
\end{equation}
where $\nu_f = \V_\pi(f) = \E_\pi \{ f^2(\omega) \} - \E_\pi^2 \{ f(\omega) \}$ is the
variance of $f$ with respect to $\pi$, and $\rho_j = \mathrm{corr}( f(\omega^{(t)}),
f(\omega^{(t + j)}) )$ is the lag-$j$ autocorrelation \citep{Peskun1973}. The
\emph{efficiency} of the MCMC for estimating $I$ is defined as the variance ratio
\begin{equation} \label{eq:eff}
   E = \frac{\nu_f}{\nu} = \frac{1}{1+2\sum_{j=1}^\infty \rho_j}.
\end{equation}
The product $NE$ is also known as the \textit{effective sample size} (ESS).  Note that
$E=1$ for an independent sampler, which draws $\omega$ directly from the target
$\pi(\omega)$. We compute $E$ using the initial positive sequence method and the batch
method \citep{Geyer1992}.

The general problem we pose in this paper is the following: given the target distribution
$\pi(k, \theta_k)$ and the function $f$, what is the Markov chain $P$ that minimizes $\nu$
or maximizes $E$?  We focus on the use of the MCMC sample to estimate the posterior
probability for a particular model, such as the MAP model.  Note that to estimate $\pi_1$,
we set $f(k, \theta_k) = 1$ if $k = 1$ and $0$ otherwise.

\section{The case of two models with and without parameters}

\subsection{Background}

The case of two models with no parameters ($H_0$ and $H_1$) has been well-studied
\cite[e.g.,][]{Frigessi1992}.  The unique transition matrix that achieves both the maximum
$\Pj$ and maximum efficiency, i.e.,
\begin{align} \label{eq:K2PjEbest}
  \Pj^* &= 2\min\{\pi_0, \pi_1\},              \nonumber \\
  E^*                 &= \frac{1}{1 - 2\min\{\pi_0, \pi_1\}},
\end{align}
is
$P =
\Bigl[
\begin{smallmatrix}
  0                   & 1 \\
  \frac{\pi_0}{\pi_1} & 1-\frac{\pi_0}{\pi_1}
\end{smallmatrix}
\Bigr]$
if $\pi_0 \le \frac{1}{2}$, and
$P =
\Bigl[
\begin{smallmatrix}
  1-\frac{\pi_1}{\pi_0} & \frac{\pi_1}{\pi_0} \\
  1                     & 0
\end{smallmatrix}
\Bigr]$ %
if $\pi_0 > \frac{1}{2}$.  In both cases, the optimal chain leaves the less probable model
immediately.

The case of two models with parameters is more complex.  In general the sampled model
index $k$ may not constitute a Markov chain.  $\Pj$ of eq.~\ref{eq:K2PjEbest} is still the
maximum, as in the case of no parameters (see SI text 1).  Here we consider the case in
which there are no free parameters in $H_0$ while $H_1$ has parameters.  Let the target
distribution be $\pi_0$ and $\pi_1 = 1 - \pi_0$, with $\theta \sim \pi_1(\theta)$ in
$H_1$.  We consider two algorithms.  In algorithm A0 (no within-model move), each MCMC
iteration consists of only one cross-model move, with $q_{01} = q_{10} = 1$, and
$q(\theta|k=0,k'=1) = g(\theta)$ in the $0 \to 1$ move.  In algorithm A1 (with
within-model move), we in addition include a within-model step in each MCMC iteration,
which means sampling $\theta$ from the target $\pi_1(\theta)$ if the chain is in $H_1$. 
Note that if $g(\cdot)$ and $\pi_1(\cdot)$ have the same support, the cross-model move
alone will give a correct algorithm, and a within-model move is not essential.  Otherwise
if the support of $g(\cdot)$ is a subset of $\Theta_1$, within-model moves that change
$\theta$ in $H_1$ will be necessary.  The $0\to 1$ move is accepted with probability
$\min\{1, A_{01}\}$, where
\begin{equation}
   A_{01} = \tfrac{1}{g(\theta)} \times \tfrac{\pi_1 \pi_1(\theta)}{\pi_0}.
   \label{eq:A01}
\end{equation}

In the reverse $1\to 0$ move, we simply drop $\theta$.  The move is accepted
with probability $\min\{1, A_{10}\}$, with $A_{10} = \frac{1}{A_{01}}$.

The model-jump probability defined in eq.~\ref{eq:pjump} is
\begin{align}
   \Pj &= \pi_0\int_{-\infty}^\infty g(\theta) \min\{1,A_{01}\}\d \theta   \nonumber \\
   &\phantom{==} + (1 - \pi_0)\int_{-\infty}^\infty \pi_1(\theta) \min\{1, A_{10}\} \d \theta  \nonumber \\
   &= 2\pi_0\int_{-\infty}^\infty g(\theta) \min\{1,A_{01}\}\d \theta.
   \label{eq:pjumpRJ}
\end{align}

We define the entrance and exit distributions when the Markov chain moves from model $i$
to model $j$.

Definition 1.  The \emph{entrance distribution} $f_{ij}^\text{in}(\theta_j)$ for parameter
$\theta_j$ in model $H_j$ is the distribution of $\theta_j$ immediately after the
stationary $\pi$-reversible Markov chain moves from model $i$ to model $j$.  Similarly we
define the \emph{exit distribution} $f_{ji}^\text{out}(\theta_j)$ as the distribution of
$\theta_j$ immediately before the chain moves from model $j$ into model $i$.

Due to detailed balance, $f_{ij}^\text{in}(\theta_j) = f_{ji}^\text{out}(\theta_j)$ for
any $i \ne j$, so that the entrance and exit distributions are equivalent.  But both may
differ from the target $\pi(\theta_j | j)$.

In the case considered here, the entrance distribution is
\begin{equation} \label{eq:entrance}
   f_{01}^\text{in}(\theta) =
   \begin{cases}
      \frac{1}{z} g(\theta), & \text{if } A_{01}(\theta) \ge 1, \\
      \frac{1}{z} \frac{\pi_1}{\pi_0}\pi_1(\theta), & \text{otherwise},
   \end{cases}
\end{equation}
where $z = \int_{-\infty}^\infty g(\theta) \min\{1,A_{01}\}\d \theta = \Pj/(2\pi_0)$.

We analyze two examples with the uniform and normal target distributions, respectively.
These are interesting as any target distribution can be transformed into the uniform, and
in large datasets the posterior is approximately normal.

\subsection{Uniform distribution example}

Consider two models $H_0, H_1$ with the target $\pi_0, \pi_1 = 1-\pi_0$, and
$\pi(\theta|1) = 1$, $0 \le \theta \le 1$.  The A0 algorithm consists of a cross-model
move in each MCMC iteration.  In the $0\to 1$ move, we propose $\theta$ from a symmetric
beta distribution $\theta \sim \text{beta}(a, a)$, with density $g(\theta) =
\frac{1}{B(a,a)}\theta^{a - 1}{(1 - \theta)}^{a - 1}$, $0 \le \theta \le1$, where $B(a,b)$
is the beta function.  This has the $\cup$, \textemdash, and $\cap$ shapes, for $a < 1$,
$= 1$, and $ > 1$, respectively.  The proposal is accepted with probability
\begin{equation}
   \alpha_{01} = \min\{1, A_{01}\}
   = \min\bigl\{1, \tfrac{1}{g(\theta)} \cdot \tfrac{\pi_1}{\pi_0} \bigr\}.
   \label{eq:A01U}
\end{equation}
In the reverse $1\to 0$ move, we simply drop $\theta$.  The move is accepted with
probability $\alpha_{10} = \min\{1, A_{10}\}$, where $A_{10} = \frac{1}{A_{01}}$.

The model-jump probability (eq.~\ref{eq:pjumpRJ}) is 
\begin{equation}
   \Pj = 2\pi_0 \int_0^1 g(\theta)
   \min\Bigl\{1, \tfrac{1}{g(\theta)}\tfrac{\pi_1}{\pi_0} \Bigr\} \d \theta.
   \label{eq:pjumpU}
\end{equation}

In the A1 algorithm, we include a within-model move, sampling $\theta$ from the target
$\U(0, 1)$, if the chain is in $H_1$.

\begin{figure*}[t]
   \centering
   \includegraphics[scale=0.75]{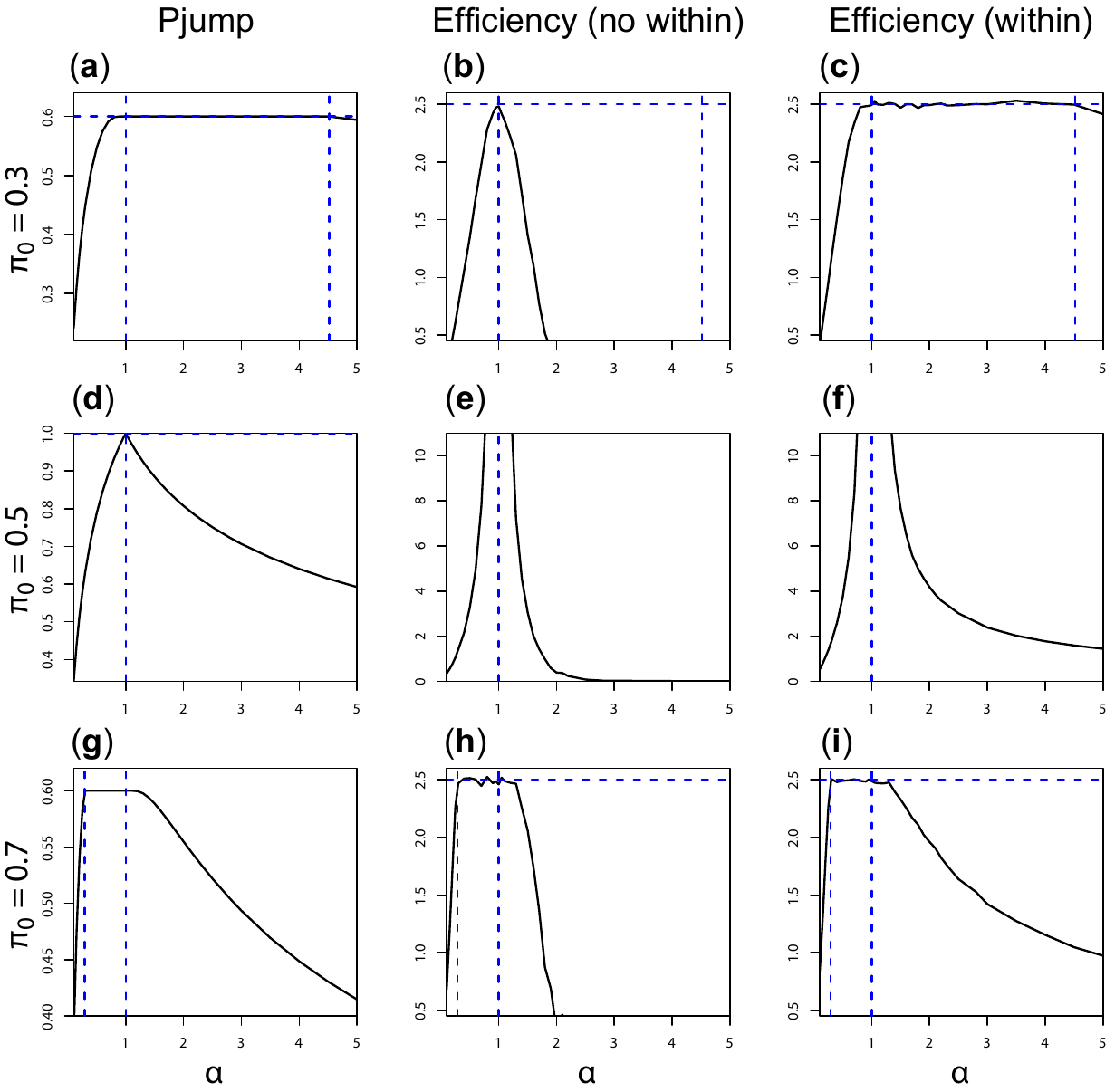}
   
   \caption{$\Pj$ and $E$ plotted as functions of $a$ in the proposal density beta$(a, a)$
      for the Uniform-Distribution example.  The three rows correspond to $\pi_0=0.3$, $0.5$
      and $0.7$.  The vertical blue dashed lines indicate the range of $a$ in which $\Pj$ is
      at the maximum: $[1, 4.51879]$ for $\pi_0 = 0.3$, and $[0.29058, 1]$ for $\pi_0 = 0.7$
      (eq.~\ref{eq:uniform-astar}). $\Pj$ and $E$ are calculated by running the MCMC algorithm
      over $10^8$ iterations.  \\ %
      (\textbf{a}-\textbf{c}) When $\pi_0 = 0.3$, the maximum $\Pj^* = 2\pi_0 = 0.6$ is
      achieved in the range $1 \le a \le a^*$ with $a^* = 4.51879$.  For all MCMC algorithms
      indexed by $a \in (1,a^*]$, every move from $H_0$ to $H_1$ is accepted, with $A_{01}\ge
      1$ (eq.~\ref{eq:A01U}) and $g(\cdot) = f_{01}^\text{in}(\cdot) \ne \pi_1(\cdot)$, but
      without the within-model move, $E$ is not optimal except at $a = 1$ (\textbf{b}).  This
      is because even though $\rho_1$ achieves the value for the optimal chain, $\rho_k$,
      $k\ge2$, are not optimal (see SI text 2 Autocorrelation functions).  
      (\textbf{c}) Including a within-model move to restore the parameter to its stationary
      distribution leads to the maximum $E^* = 2.5$ for the whole range $1 \le a \le a^*$. 
      All those optimal algorithms have the same Markov chain transition kernel. \\ %
      (\textbf{d}--\textbf{f}) When $\pi_0 = 0.5$, the maximum $\Pj = 100\%$ and the maximum
      $E^* = \infty$ are achieved at one point ($a=1$), at which $g(\cdot) =
      f_{01}^\text{in}(\cdot) = \pi_1(\cdot)$.  \\ %
      (\textbf{g}-\textbf{i}) When $\pi_0 = 0.7$, the maximum $\Pj^* = 2\pi_1 = 0.6$ is
      achieved in the range $a^* \le a \le 1$ with $a^* = 0.29058$.  In this interval
      $A_{01}\le 1$ for all parameter values, so that $\theta$ has the correct stationary
      distribution as soon as the chain moves into $H_1$, with $g(\cdot) \ne
      f_{01}^\text{in}(\cdot) = \pi_1(\cdot)$.  As a result, algorithm A0 (with no
      within-model move) achieves the maximum $E^* = 2.5$ for the whole range $a^* \le a \le
      1$ (\textbf{h}).  All proposal algorithms specified by $a \in [a^*, 1]$ lead to the same
      Markov chain. %
      (\textbf{i}) Including a within-model move does not improve $E$ any further when $a^*
      \le a \le 1$. \\ %
   }  \label{fig:uniform-pjump-eff}
\end{figure*}

\begin{figure*}[th]
   \centering
   \includegraphics[scale=0.75]{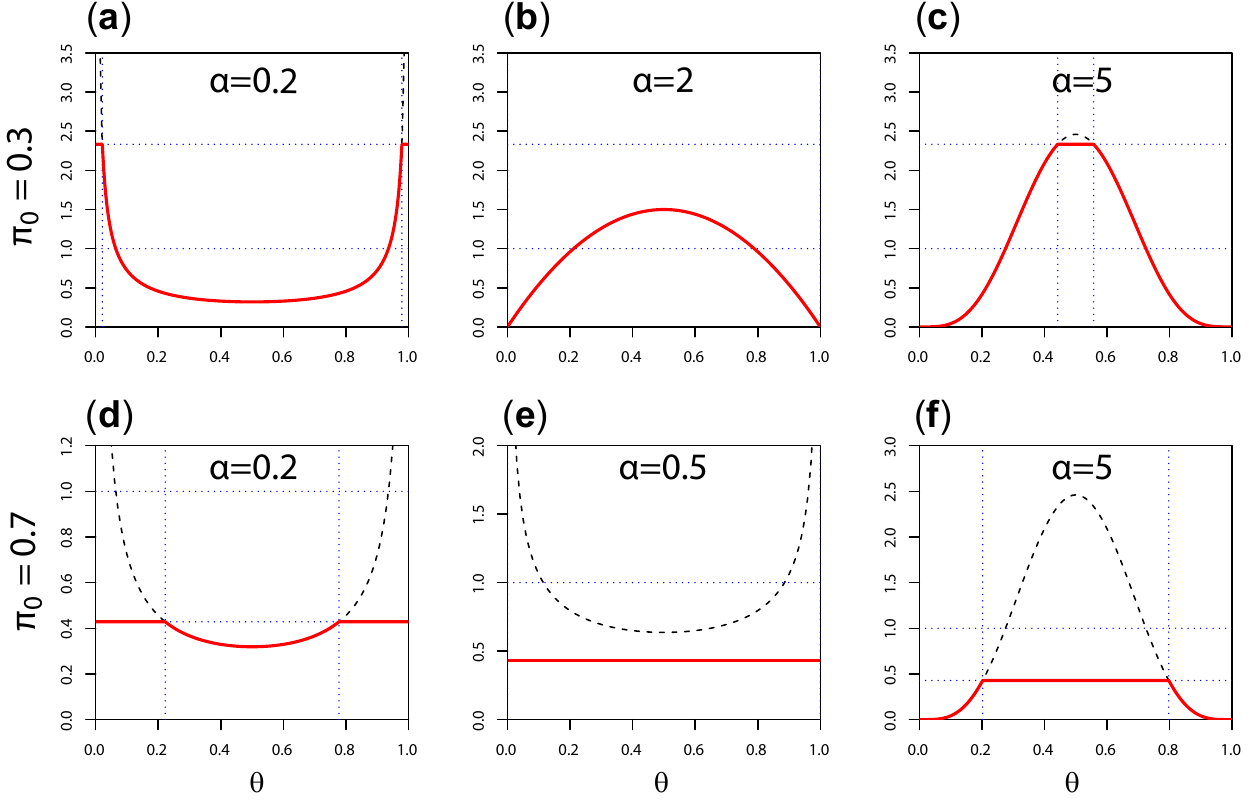}
   
   \caption{The proposal distribution (dashed line) and (unnormalized) entrance
      distribution (red solid line) of $\theta$ in the Uniform-Distribution example.
      Horizontal dotted lines indicate 1 and $\pi_1/\pi_0$.  Vertical dotted lines indicate
      $\theta_1^*$ and $\theta_2^*$, the solutions to eq.~\ref{eq:uniform-theta12}. \\ %
      (\textbf{a}) $\pi_0 = 0.3, a = 0.2: \theta_1^* = 0.0212, \theta_2^* = 0.9788$, $\Pj <
      \Pj^*, E < E^*$. \\ %
      (\textbf{b}) $\pi_0 = 0.3, a = 2$: $g(\cdot) = f_{01}^\text{in}(\cdot) \ne
      \pi_1(\cdot)$, $\Pj = \Pj^*$ but $E < E^*$. \\ %
      (\textbf{c}) $\pi_0 = 0.3, a = 5:   \theta_1^* = 0.4425, \theta_2^* = 0.5575$, $\Pj <
      \Pj^*, E < E^*$.  \\ %
      (\textbf{d}) $\pi_0 = 0.7, a = 0.2: \theta_1^* = 0.2223, \theta_2^* = 0.7777$, $\Pj <
      \Pj^*, E < E^*$.  \\ %
      (\textbf{e}) $\pi_0 = 0.7, a = 0.5$: $g(\cdot) \ne f_{01}^\text{in}(\cdot) =
      \pi_1(\cdot)$, $\Pj = \Pj^*$, $E = E^*$. \\ %
      (\textbf{f}) $\pi_0 = 0.7, a = 5:   \theta_1^* = 0.2025, \theta_2^* = 0.7975$, $\Pj <
      \Pj^*, E < E^*$. \\ %
      In cases \textbf{a}, \textbf{c}, \textbf{d}, and \textbf{f}, the proposal distribution
      $g(\theta)$, the entrance distribution $f_{01}^\text{in}(\theta)$, and the target
      distribution $\pi_1(\theta)$ are all different, so that algorithm A0 achieves neither
      $\Pj^*$ nor $E^*$.  In case \textbf{b}, the A0 algorithm achieves $\Pj^*$ but not $E^*$.
      In case \textbf{e}, algorithm A0 achieves both $\Pj^*$ and $E^*$. \\ %
   } \label{fig:uniform-entrance}
\end{figure*}

\textit{\bf (i) The case of $\pi_0 \le \frac{1}{2}$} (e.g., $\pi_0=0.3$ in
figs.~\ref{fig:uniform-pjump-eff}\&\ref{fig:uniform-entrance}).  In this case, the
maximum $\Pj$, at $2\pi_0$, is achieved by having $A_{01} \ge 1$, or
equivalently, $g(\theta) \le \frac{\pi_1}{\pi_0}$  for all $\theta$.  This is possible if
and only if $a \ge 1$ and $g(\frac{1}{2}) \le \frac{\pi_1}{\pi_0}$, or equivalently, $1
\le a \le a^*$, where $a^*$ is the unique solution to the equation
\begin{equation}  \label{eq:uniform-astar}
   g \bigl( \tfrac{1}{2} \bigr) = \tfrac{1}{B(a, a)} \cdot \bigl( \tfrac{1}{4} \bigr)^{a - 1} = \tfrac{\pi_1}{\pi_0}.
\end{equation}
When $\pi_0 = 0.3$ (fig.~\ref{fig:uniform-pjump-eff}\textbf{a}\&\textbf{c}),
eq.~\ref{eq:uniform-astar} gives $a^* = 4.51879$.  The case of $a = 2 \in [1, a^*]$ is
shown in figure \ref{fig:uniform-entrance}\textbf{b}.

\begin{figure*}[t]
   \centering
   \includegraphics[scale=0.80]{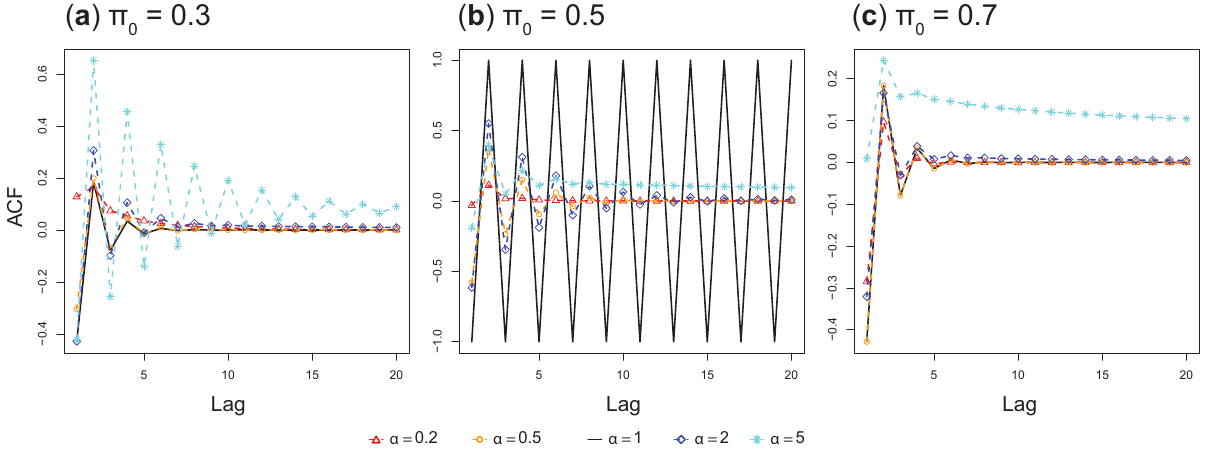}
   
   \caption{Autocorrelation function for the model index in the Uniform-Distribution
      example for different proposal distributions beta($a, a$), with $a = 0.2, 0.5, 1, 2$,
      and 5.  When $\pi_0 = 0.3$, $\rho_1$ is the same between $a = 1$ and 2, as with both
      $a$ values the chain achieves the maximum $\Pj$, but $\rho_k$ ($k > 1$) are different
      between the two $a$ values.  When $\pi_0 = 0.7$, $\rho_k$ for all $k \ge 1$ are the
      same between $a = 1$ and 0.5, and with both $a$ values the chain achieves the maximum
      $\Pj$ and maximum $E$. %
   } \label{fig:uniform-ACF}
\end{figure*}

In this case ($\pi_0 < \frac{1}{2}$ and $1 \le a \le a^*)$, the entrance distribution
matches the proposal, with $f^{\text{in}}_{01}(\theta) = g(\theta)$ for all $\theta$, but
both differ from the target $\pi_1(\theta)$.  To see that algorithm A0 (without the
within-model move) still samples correctly from the posterior, note that the probability
of leaving model $H_1$ depends on $\theta$, as $A_{10} < 1$ for all $\theta$.  Each value
of $a$ may be considered an MCMC algorithm, consisting of one cross-model move with
$q_{01} = q_{10} = 1, q(\theta|0, 1) = g(\theta)$.  Thus for the whole class of A0
algorithms indexed by $a \in [1, a^*]$, $\Pj$ is at its maximum but $E$ is not except for
$a=1$.  When $1 < a \le a^*$, the autocorrelation $\rho_1$ is at its minimum (note that
with two states, $\rho_1$ is a simple function of $\Pj$), but $\rho_k$, $k \ge 2$ are not
(fig.~\ref{fig:uniform-ACF}), which explains why $E$ does not reach the optimum.  This is
analyzed in SI text 2.  If we add a within-model move sampling $\theta$ from its target
$\pi_1(\theta)$, all the A1 algorithms in the class with $1 \le a \le a^*$ will be
optimal, achieving $E^*$, and they all specify the same Markov chain transition kernel.

When $a$ is outside the interval $[1, a^*]$, $A_{01} > 1$ for some values of $\theta$ and
$<1$ for others.  The entrance distribution is continuous but not smooth (see
fig.~\ref{fig:uniform-entrance}\textbf{a}\&\textbf{c} in the case of $\pi_0=0.3$).  The
proposal $g(\cdot)$, the entrance $f_{01}^\text{in}(\cdot)$, and the target $\pi_1(\cdot)$
are all different. The value of $\theta$ at which $A_{01} = 1$ is given by the quadratic
equation
\begin{equation} \label{eq:uniform-theta12}
   \theta^2 - \theta + c = 0,
\end{equation}
where $ c = {\bigl[\frac{\pi_1}{\pi_0} B(a,a)\bigr]}^{1/(a-1)} $.  The two solutions are
$\theta_{1,2}^* = \frac{1}{2} \mp \frac{1}{2}\sqrt{1 - 4c}$.

Thus if $\pi_0 \le \frac{1}{2}$ and $a < 1$ (e.g., fig.~\ref{fig:uniform-entrance}a,
$\pi_0=0.3, a=0.2$)
\begin{align} \label{eq:uniform-Pj-a}
   \Pj &= 2\pi_0 \times 2\int_0^{\theta_1^*} \frac{\pi_1}{\pi_0} \d\theta
                      + 2\pi_0 \int_{\theta_1^*}^{\theta_2^*} g(\theta) \d \theta  \nonumber \\
                     &= 4\pi_1\theta_1^* + 2\pi_0 [G(\theta_2^*) - G(\theta_1^*) ],
\end{align}
where $G(\theta) = \int_0^\theta g(\xi) \d \xi$ is the cumulative density function (CDF)
of the proposal distribution.  The entrance distribution when the chain moves from 0 to 1
is
\begin{equation} \label{eq:uniform-entrancea}
   f_{01}^\text{in}(\theta) =
   \begin{cases}
      \frac{g(\theta)}{z},  & \text{if } \theta \in [\theta_1^*, \theta_2^*], \\
      \frac{\pi_1/\pi_0}{z},& \text{otherwise},
   \end{cases}
\end{equation}
where $z=\Pj/(2\pi_0)=2\frac{\pi_1}{\pi_0}\theta_1^* + [G(\theta_2^*) - G(\theta_1^*) ]$. 

If $\pi_0 \le \frac{1}{2}$ and $a > a^*$ (e.g., fig.~\ref{fig:uniform-entrance}c,
$\pi_0=0.3, a=5$),
\begin{align} \label{eq:uniform-Pj-b}
   \Pj &= 2\pi_0 \times 2 \int_0^{\theta_1^*} g(\theta) \d \theta
                      + 2\pi_0 \int_{\theta_1^*}^{\theta_2^*} \frac{\pi_1}{\pi_0} \d\theta  \nonumber \\
       &= 4\pi_0 G(\theta_1^*) + 2\pi_1(\theta_2^* - \theta_1^*),
\end{align}
and the entrance distribution is
\begin{equation} \label{eq:uniform-entrance-b}
   f_{01}^\text{in}(\theta) =
   \begin{cases}
      \frac{\pi_1/\pi_0}{z}, & \text{if } \theta \in [\theta_1^*, \theta_2^*], \\
      \frac{g(\theta)}{z},   & \text{otherwise},
   \end{cases}
\end{equation}
where $z = 2 G(\theta_1^*)+\frac{\pi_1}{\pi_0}(\theta_2^* - \theta_1^*)$.

\textit{\bf (ii) The case of $\pi_0 > \frac{1}{2}$} (e.g., $\pi_0=0.7$ in
figs.~\ref{fig:uniform-pjump-eff}\&\ref{fig:uniform-entrance}).  The maximum $\Pj^* =
2\pi_1$ is achieved by having $A_{01} \le 1$, or equivalently $g(\theta) \ge
\frac{\pi_1}{\pi_0}$, for all $\theta$.  This is possible if and only if $a \le 1$ and
$g(\frac{1}{2}) \ge \frac{\pi_1}{\pi_0}$, or equivalently if and only if $a^* \le a \le
1$, where $a^*$ is the solution to eq.~\ref{eq:uniform-astar}.  In the example of figure
\ref{fig:uniform-pjump-eff} ($\pi_0 = 0.7$), eq.~\ref{eq:uniform-astar} gives $a^* =
0.29058$.  The value $a=0.5$ (fig.~\ref{fig:uniform-entrance}\textbf{e}, $\pi_0 = 0.7$) is
in the interval $[a^*,1]$.

In this case ($\pi_0 > \frac{1}{2}$ and $a^* \le a \le 1$ so that $A_{01} \le 1$ for all
$\theta$), the entrance distribution matches the target, $f_{01}^\text{in}(\theta) =
\pi_1(\theta)$ for all $\theta$, and both $\Pj$ and $E$ reach the maximum.  It is easy to
see that the whole class of A0 algorithms based on the proposal beta$(a,a)$ with $a^* \le
a \le 1$ generate the same Markov chain.  In other words, the most efficient Markov chain
is unique but it can be generated by a whole class of MCMC proposal algorithms.

When $a$ is outside the interval $[a^*,  1]$, $A_{01}>1$ for some values of $\theta$ and
$<1$ for others.  The value of $\theta$ at which $A_{01} = 1$ is given by
eq.~\ref{eq:uniform-theta12}.  Thus if $\pi_0 > \frac{1}{2}$ and $a < a^*$, $\Pj$ is
given by eq.~\ref{eq:uniform-Pj-a} and the entrance distribution by
eq.~\ref{eq:uniform-entrancea} (see, e.g., fig.~\ref{fig:uniform-entrance}\textbf{d} for
$\pi_0=0.7, a=0.2$).  If $\pi_0 > \frac{1}{2}$ and $a>1$, $\Pj$ is given by
eq.~\ref{eq:uniform-Pj-b} and the entrance distribution by
eq.~\ref{eq:uniform-entrance-b} (see, e.g., fig.~\ref{fig:uniform-entrance}\textbf{f},
$\pi_0=0.7, a=5$).

Finally if $\pi_0 = \frac{1}{2}$, both $\Pj$ and $E$ reach their maxima at one point, $a =
1$, in which case the proposal is the target $g(\cdot) = \pi_1(\cdot)$, and the maximum $E
\to \infty$ (Fig.~\ref{fig:uniform-pjump-eff}\textbf{d}--\textbf{f}).  In this case the
chain degenerates into a periodic chain.  One can introduce a small perturbation, with
$p_{00} = \epsilon$ and $p_{01} = 1 - \epsilon$, say, to remove the periodicity.

\subsection{Gaussian-distribution example}

Suppose $H_0$ has no free parameters and $H_1$ has one parameter $\mu$, with the posterior
$\pi_0$ and $\pi_1 = 1 - \pi_0$ for the two models, and $\mu \sim \N(0, 1)$ within $H_1$. 
We use $g(\mu) = \phi(\mu; \mu_p, \sigma_p^2)$ as the proposal density when the chain
moves from 0 to 1, where $\phi(z; \mu, \sigma^2)$ is the probability density function
(PDF) for $\N(\mu, \sigma^2)$.  Algorithm A0 thus consists of one cross-model move with
$q_{01} = q_{10} = 1$ and $q(\mu|0,1) = \phi(\mu; \mu_p, \sigma_p^2)$.  Algorithm A1
includes a within-model step, which is a random draw from the target $\mu \sim \N(0, 1)$
when the chain is in 1.

We derive $\Pj$ and $E$ for different proposal algorithms $(\mu_p, \sigma_p^2)$ in SI text
3.  Example calculations are shown in fig.~\ref{fig:normal-pjump-eff}.  When $\pi_0 \le
\frac{1}{2}$ (e.g., $\pi_0=0.3$), the maximum $\Pj^* = 2\pi_0$ is achieved by a class of
A0 algorithms represented by the round flat top of the table mountain in
fig.~\ref{fig:normal-pjump-eff}\textbf{a}, given by eq.~\ref{eq:N-round-table-pi<0.5}. In
those algorithms, every $0\to 1$ proposal is accepted as $A_{01} \ge 1$ for all $\mu$, and
the proposal and entrance distributions match, but both differ from the posterior,
$g(\cdot) = f_{01}^\text{in}(\cdot) \ne \pi_1(\cdot)$.  Thus $\Pj^*$ is achieved, but $E$
may not be optimal except when $g(\cdot) = \pi_1(\cdot)$.  Including the within-model move
in the A1 algorithm to restore $\mu$ to its stationary distribution achieves $E^*$ for the
class of algorithms.  When the proposal is outside the flat top, algorithm A1 including
the within-model move has higher efficiency $E$ than algorithm A0 without the within-model
move, but neither algorithm achieves the maximum $\Pj$ or maximum $E$.

The case of $\pi_0 > \frac{1}{2}$ is represented by $\pi_0 = 0.7$ in figure
\ref{fig:normal-pjump-eff}\textbf{g}--\textbf{i}.  Maximum $\Pj^* = 0.6$ and maximum $E^*
= 2.5$ are achieved when $(\mu_p, \sigma_p)$ are in a region given by
eq.~\ref{eq:N-round-table-pi>0.5}, represented by the flat top of the table mountain in
figure \ref{fig:normal-pjump-eff}\textbf{g}--\textbf{i}.  In all such algorithms, $A_{01}
\le 1$ for all $\mu$ and as a result, algorithm A0 without the within-model move already
achieves $E^*$, so that including a within-model move in A1 does not have an effect. When
the proposal is outside the flat top, algorithms A0 and A1 do not achieve $\Pj^*$ or
$E^*$, and including within-model moves in A1 improves the mixing efficiency $E$.

\subsection{General case of the Gaussian example}

Suppose the observed data are $X = \{x_1,\dots,x_n\}$, consisting of $n$ observations with
the sample mean $\bar{x} = \frac{1}{n}\sum_{i=1}^n x_i$.  We consider two candidate
models: $H_0$: $\N(0, \sigma^2)$ and $H_1$: $\N(\mu, \sigma^2)$, with $\sigma^2$ known.
This is a standard problem of hypothesis testing.  We assign the prior probability
$\frac{1}{2}$ to each model and the prior $\mu \sim \N(\mu_0,\sigma_0^2)$ for $\mu$ in
$H_1$.

The likelihood function is
\begin{equation} \label{eq:normal-L0}
   L_0 = \prod_{i=1}^n \phi(x_i;0,\sigma^2) \propto\frac{1}{\sqrt{2\pi\sigma^2/n}} \mathrm{e}^{-\frac{n}{2\sigma^2}\bar{x}^2}
   = \phi\bigl(\bar{x}; 0, \tfrac{\sigma^2}{n}\bigr)
\end{equation}
under model $H_0$ and 
\begin{equation} \label{eq:normal-L1}
   L_1(\mu) = \prod_{i=1}^n \phi(x_i;\mu,\sigma^2)
           \propto 
           \phi\bigl(\bar x; \mu, \tfrac{\sigma^2}{n} \bigr)
\end{equation}
under $H_1$.

Within $H_1$, we have $p(\mu) p(\bar{x}|\mu) = p(\bar{x}) p(\mu|\bar{x})$, or
\begin{multline}
  \phi\bigl( \mu; \mu_0,\sigma_0^2 \bigr)
  \phi\bigl(\bar{x};\mu,\tfrac{\sigma^2}{n}\bigr)  \\
  =  \phi\bigl(\bar{x};\mu_0, \tfrac{\sigma^2}{n} + \sigma_0^2\bigr)
  \phi\bigl( \mu; \tfrac{n\bar{x}/\sigma^2 + \mu_0/\sigma_0^2}{n/\sigma^2+1/\sigma_0^2}, \tfrac{1}{n/\sigma^2+1/\sigma_0^2} \bigr) ,
\end{multline}
where $\frac{1}{\sigma_0^2}$ is the prior precision, $\frac{n}{\sigma^2}$ is the sample
precision, while the posterior precision $\frac{1}{\sigma_1^2} = \frac{n}{\sigma^2} +
1/\sigma_0^2$ is the sum of the two.  The posterior mean $\mu_1 = \frac{n\bar{x}/\sigma^2
   + \mu_0/\sigma_0^2}{n/\sigma^2 + 1/\sigma_0^2}$ is the weighted average of the prior mean
and the sample mean.

Furthermore,  the marginal likelihood for $H_1$ is $M_1 = \phi \bigl( \bar{x}; \mu_0,
\frac{\sigma^2}{n} + \sigma_0^2 \bigr)$.  The posterior probability for $H_0$ is
thus
\begin{equation} \label{eq:normal-pi0}
   \pi_0 = \frac{\frac{1}{2}L_0}{\frac{1}{2}L_0 + \frac{1}{2} M_1}
         = \frac{1}
                {1 + \frac{\sigma_1}{\sigma_0}
                	\exp \biggl\{ \frac{1}{2} \Bigl(\frac{\mu_1^2}{\sigma_1^2} 
                                               - \frac{\mu_0^2}{\sigma_0^2}
                                           \Bigr)
                       \biggr\}
                } .
\end{equation}

In the MCMC algorithm, we propose $\mu$ from $\N(m_p, s_p^2)$ to move from $H_0$ to $H_1$.
Define $\mu_p = (m_p - \mu_1)/\sigma_1$ and $\sigma_p = s_p/\sigma_1$.  Then our theory
concerning $\Pj$ and $E$ in SI text 3 applies.

We use this theory to illustrate the common observation that when the datasize increases,
trans-model algorithms tend to become less and less efficient because the proposal for
parameters in the new model is not adapted to closely match the posterior.  When the
datasize increases, the within-model posterior becomes more concentrated, and the proposed
values of parameters for the new model are increasingly likely to miss the mode of the
posterior, leading to rejection of the proposed model even if it has a higher posterior
than the current model. %

\begin{figure*} [t]
   \centering %
   \includegraphics[scale=0.75]{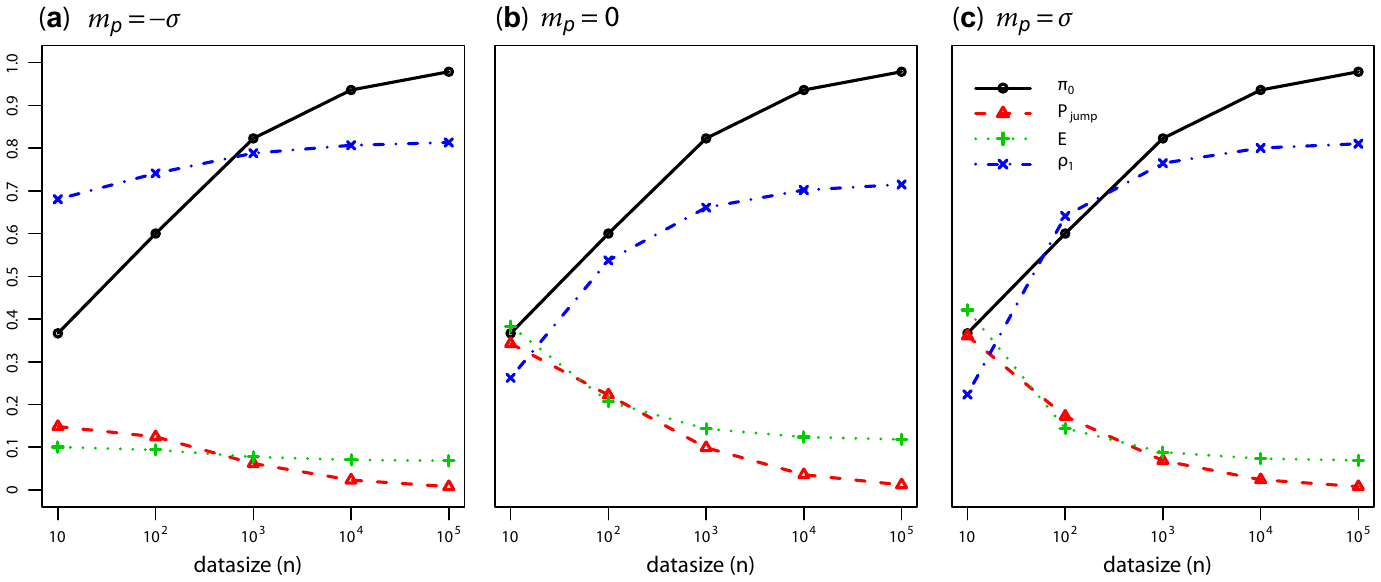} %
   
   \caption{Model-jump probability ($\Pj$) and efficiency ($E$) plotted against the
      datasize ($n$) in the trans-model MCMC algorithm comparing two models: $H_0$: $\N(0,
      \sigma^2)$ and $H_1$: $\N(\mu, \sigma^2)$, with $\sigma^2$ assumed known and
      $\sqrt{n}\bar x = 1.96\sigma$ fixed.  When $n$ increases, the fixed proposal (with
      $m_s = -0.1\sigma, 0, 0.1\sigma$ and $s_p = \sigma^2$) becomes farther away from the
      posterior and both $\Pj$ and $E$ deteriorate. Shown also are the posterior model
      probability $\pi_0 = \P\{H_0|\bar x\}$ and the lag-1 autocorrelation for model index
      ($\rho_1$), with smaller $\rho_1$ for more efficient algorithms. \\ %
   } \label{fig:normal-pjump-eff-n}
\end{figure*}

We fix $\sqrt{n}\bar x = 1.96\sigma$ so that the $p$-value for the likelihood ratio test
(LRT) remains 5\% but let the sample size $n$ increase.  We use the prior $\mu\sim
\N(\mu_0, \sigma_0^2)$ with $\mu_0 = 0$ and $\sigma_0^2 = \sigma^2$.  Let $m_s =
-0.1\sigma, 0, 0.1\sigma$, and $s_p = \sigma$ in the proposal during the $0\to 1$ move.
Note that the proposal is fixed, independently of the posterior for $\mu$ in $H_1$.  When
the datasize $n$ increases, the proposal $g(\cdot)$ becomes increasingly further away from
the posterior mode ($\mu_1$) and also more diffuse relative to the posterior, and we see a
deterioration in both $\Pj$ and $E$ (fig.~\ref{fig:normal-pjump-eff-n}).  This is in
contrast with within-model algorithms, the performance of which is independent of the
posterior variance when the appropriate scale is chosen automatically.

Note that the $p$-value for the LRT is 5\% for all values of $n$ in
fig.~\ref{fig:normal-pjump-eff-n} but the posterior $\pi_0 = \P\{H_0|\bar x\}$ is nearly
100\% for large $n$.  This is a case of Lindley's paradox \citep{Lindley1957}, in which LRT
and Bayesian model selection reach opposite conclusions for the same data.

\section{Model-proposal probability}

Both the uniform and normal examples analyzed above involve only two models. For $K=2$
states, maximum $\Pj$ is a necessary but not sufficient condition for maximum $E$.  Here
we consider the case of $K > 2$ models.  First we focus on the case where there are no
parameters in the model, and the MCMC algorithm is specified by the model-proposal
probabilities, $Q = \{ q_{kk'} \}$.  We focus on the efficiency for estimating $\pi_1
\equiv \P(H_1|X)$.

Efficient Markov chains in the discrete case were analyzed by \citep{Frigessi1992}, and
the theory also applies when the parameters for the new model are proposed from the
posterior during the cross-model move, that is, if $q(\theta_{k'} | k, \theta_k, k') =
\pi(\theta_{k'}|k') \equiv p(\theta_{k'}|k',X)$.  We provide an overview of the theory
in \textbf{SI text 4: Super-efficient Markov chains in the discrete case}.

If $\pi_1 \le \frac{1}{2}$, the optimal Markov chain for estimating $\pi_1$ is
\begin{equation} \label{eq:P-pi<1/2}
   P^* =
   \begin{bmatrix}
      0                     & \frac{\pi_2}{1-\pi_1}             & \cdots & \frac{\pi_K}{1-\pi_1} \\
      \frac{\pi_1}{1-\pi_1} & \frac{1-2\pi_1}{(1-\pi_1)^2}\pi_2 & \cdots & \frac{1-2\pi_1}{(1-\pi_1)^2}\pi_K \\
      \vdots                & \vdots                            & \ddots & \vdots \\
      \frac{\pi_1}{1-\pi_1} & \frac{1-2\pi_1}{(1-\pi_1)^2}\pi_2 & \cdots & \frac{1-2\pi_1}{(1-\pi_1)^2}\pi_K
   \end{bmatrix}
   .
\end{equation}
This achieves the maximum $E^* = 1/(1 - 2\pi_1)$, with $\Pj = 1 -
\tfrac{1-2\pi_1}{(1-\pi_1)^2} \sum_{k\ge 2} \pi_k^2 < 1$, even though $\Pj=100\%$ is
easily achievable when $\pi_k < \frac{1}{2}$ for all $k$ and when $K>2$.
Eq.~\ref{eq:P-pi<1/2} suggests that when the Markov chain is in 1, it moves away
immediately, to other states in proportion to their probabilities.  When not in 1, it
visits the other states according to their posterior and it is worse to move too much.
Maximum $\Pj$ is neither necessary nor sufficient for achieving maximum $E$.

If $\pi_1 > \frac{1}{2}$, the optimal Markov chain for estimating $\pi_1$ is
\begin{equation} \label{eq:P-pi>1/2}
   P^* =
   \begin{bmatrix}
      1-\frac{1-\pi_1}{\pi_1} & \frac{\pi_2}{\pi_1} & \cdots & \frac{\pi_K}{\pi_1} \\
      1                       & 0                   & \cdots & 0 \\
      1                       & 0                   & \cdots & 0 \\
      \vdots                  & \vdots              & \ddots & \vdots \\
      1                       & 0                   & \cdots & 0
   \end{bmatrix}.
\end{equation}
This achieves $\Pj^* = 2(1-\pi_1)$ and $E^* =1/(2\pi_1 - 1)$.

Table \ref{tab:P3x3-example} provides a numerical example for the case of three states,
with $\pi = (\pi_1, \pi_2, \pi_3) = (\frac{4}{7}, \frac{2}{7}, \frac{1}{7})$.  The most
efficient Markov chain ($P^*$) for estimating the model probabilities is given by
eq.~\ref{eq:P-pi>1/2} for $\pi_1$ and by eq.~\ref{eq:P-pi<1/2}
for $\pi_2$ and $\pi_3$. In all three cases, the MCMC sampler $P^*$ is more efficient than
the independent sampler, with $E^* > 1$.

We note that the optimal Markov chain $P^*$ can be achieved by many proposal algorithms
($Q$). For example, $P^*$ of eq.~\ref{eq:P-pi>1/2} for $\pi_1 > \frac{1}{2}$ can be
generated by a class of proposals
\begin{equation}
  Q_a =  \{q_{ij}\} =
  \begin{bmatrix}
    1-a & \frac{a\pi_2}{1-\pi_1} & \cdots & \frac{a\pi_K}{1-\pi_1} \\
    1                          & 0                     & \cdots & 0 \\
    1                          & 0                     & \cdots & 0 \\
    \vdots                     & \vdots                & \ddots & \vdots \\
    1                          & 0                     & \cdots & 0
  \end{bmatrix},
\end{equation}
indexed by $a \in \bigl[ \frac{1-\pi_1}{\pi_1}, 1 \bigr]$.  It is easy to confirm that
this $Q$ leads to $P^*$ of eq.~\ref{eq:P-pi>1/2}.  For example, the proposal from 1 to $k
\ge 2$ is accepted with probability $\alpha_{1k} = \min\bigl\{1,
\frac{1-\pi_1}{a\pi_k}\frac{\pi_k}{\pi_1} \bigr\} = \frac{1-\pi_1}{a\pi_1}$, so that
$p_{1k} = q_{1k} \alpha_{1k} = \frac{\pi_k}{\pi_1}$. Similarly $\alpha_{k1} = 1$ and
$P_{k1} = 1$.  Thus all algorithms $Q_a$ with $\frac{1-\pi_1}{\pi_1} \le a \le 1$ are
optimal and achieve $E^*$.

Consider the two extremes, at $a = 1$ and $\frac{1-\pi_1}{\pi_1}$:
\begin{equation} \label{eq:Q1Q2}
\begin{small}
  Q_1 =
  \setlength{\arraycolsep}{1.5pt}
  \begin{bmatrix}
    0 & \frac{\pi_2}{1-\pi_1} & \cdots & \frac{\pi_K}{1-\pi_1} \\
    1                         & 0                     & \cdots & 0 \\
    1                         & 0                     & \cdots & 0 \\
    \vdots                    & \vdots                & \ddots & \vdots \\
    1                         & 0                     & \cdots & 0
  \end{bmatrix}, \hspace{0.25em}
  Q_2 =
\begin{bmatrix}
  1 - \frac{1-\pi_1}{\pi_1} & \frac{\pi_2}{\pi_1} & \cdots & \frac{\pi_K}{\pi_1} \\
  1                         & 0                     & \cdots & 0 \\
  1                         & 0                     & \cdots & 0 \\
  \vdots                    & \vdots                & \ddots & \vdots \\
  1                         & 0                     & \cdots & 0
\end{bmatrix} .
\end{small}
\end{equation}
With $Q_1$ one always attempts a model jump.  Moves into 1 are all accepted, but moves
away from 1 are accepted with probability $\alpha_{1k} = \frac{1 - \pi_1}{\pi_1}$, so
that on average a proportion $R = \sum_{k \ge 2} \pi_1 q_{1k}(1-\alpha_{1k}) = \sum_{k
   \ge 2} \pi_1 \frac{\pi_k}{1-\pi_1} \frac{2\pi_1 - 1}{\pi_1} = 2\pi_1-1$ of the
cross-model proposals are rejected.

With $Q_2$, one proposes with probability $q_{11} = 1 - \frac{1-\pi_1}{\pi_1}$, not to
attempt a move when the chain is in 1.  Then all moves are accepted with probability 1
($\alpha_{kk'} = 1$ for $k' \ne k$).  Even though $Q_1$ and $Q_2$ lead to the same Markov
chain ($P$), the computation involved may be different if computation is proportional to
the number of times $A_{kk'}$ is evaluated. In $Q_2$, when we choose not to attempt a move
when in 1, there is no need to compute $\alpha_{11}$, leading to a saving (of $2\pi_1-1 =
80\%$ for $Q_2$ relative to $Q_1$ if $\pi_1 = 0.9$). This reasoning suggests that to
minimize the computational cost, one should adjust the model-proposal probability
$q_{kk'}$ to make $A_{kk'}$ as close to 1 as possible.

\section{Optimal Markov chains in case of two models, one with parameters}

Here we extend the uniform and Guassian examples to the general case.  Consider two models
$H_0$ and $H_1$ with the posterior $\pi_0$ and $\pi_1 = 1-\pi_0$, and $\pi_1(\theta),
\theta \in (a, b)$ in $H_1$.  Again we consider algorithm A0 with no within-model move and
algorithm A1 with a within-model move.

We use $K$ bins to discretize the parameter space for $H_1$, with $a = \theta_0 < \theta_1
< \cdots < \theta_K = b$.  The resulting Markov chain will have $K+1$ states. State 1
corresponds to $H_0$ with $\pi'_1 = \pi_0$, and states $k = 2, \cdots, K+1$ to the bins
for model $H_1$, with $\pi'_k = \pi_1 \pi_1(\theta_{k-1})\Delta_{k-1}$, $k = 2, \cdots,
K+1$.  Here the differential $\Delta_{k-1} = \theta_k - \theta_{k-1} \to 0$ when $K \to
\infty$.  The dimension of $\theta$ is inconsequential; if there are two parameters in
$H_1$, $k$ can index a grid on the plane, and $\pi_1(\theta_k)\Delta_k$ will be the volume
above the $k$th grid.  The $(K+1)$-state reversible Markov chain that achieves the optimal
efficiency for estimating $\pi_0$ is given by eqs.~\ref{eq:P-pi<1/2}\&\ref{eq:P-pi>1/2}.

\subsubsection{Case $\pi_0 < \frac{1}{2}$}

In this case, eq.~\ref{eq:P-pi<1/2} gives the optimal chain for the discretized states as
\begin{equation} \label{eq:P-pi<1/2-discrete}
   P^* =
   \begin{bmatrix}
      0                   & \pi_1(\theta_1)\Delta_1     & \cdots & \pi_1(\theta_K)\Delta_K \\
      \frac{\pi_0}{\pi_1} & \frac{1-2\pi_0}{\pi_1}\pi_1(\theta_1)\Delta_1 & \cdots & \frac{1-2\pi_0}{\pi_1}\pi_1(\theta_K)\Delta_K \\
      \vdots              & \vdots                      & \ddots & \vdots \\
      \frac{\pi_0}{\pi_1} & \frac{1-2\pi_0}{\pi_1}\pi_1(\theta_1)\Delta_1 & \cdots & \frac{1-2\pi_0}{\pi_1}\pi_1(\theta_K)\Delta_K 
   \end{bmatrix}
   ,
\end{equation}
which achieves $\Pj^* = 2\pi_0$ and $E^* = 1/(1 - 2\pi_0)$.

When $K \to \infty$, the discrete chain will converge to a chain operating on the
continuous $\theta$ space, which has the transition kernel as follows.
\begin{equation} \label{eq:P-pi<1/2-kernel}
   \begin{aligned}
      &P(0 \to 0) = 0, \\
      &P(0 \to 1,\d\theta) = \pi_1(\d\theta), \\
      &P(1,\d\theta \to 0) = \tfrac{\pi_0}{\pi_1}, \\
      &P(1,\d\theta \to 1,\d\theta') = \bigl(1 - \tfrac{\pi_0}{\pi_1} \bigr) \pi_1(\d\theta').
   \end{aligned}
\end{equation}

To generate the Markov chain of eq.~\ref{eq:P-pi<1/2-kernel}, the chain must move to 1
immediately when it is in 0.  Thus we have
\begin{equation} \label{eq:pi<1/2-conditions}
   \begin{aligned}
      \text{(a) }  &{} q_{01} = 1 \text{ (always attempt model jump if in 0)};  \\
      \text{(b) }  &{} A_{01}(\theta) = \tfrac{q_{10}}{q_{01}\cdot g(\theta)} \cdot \tfrac{\pi_1 \pi_1(\theta)} {\pi_0}
      \ge 1, \ \ \text{for all } \theta ; \\
      \text{(c) }  &{} \theta \text{ should be from }\pi_1(\theta) .
   \end{aligned}
\end{equation}
Conditions (a) and (b) mean that the entrance distribution matches the proposal
distribution, and thus to satisfy condition (c), we must have $g(\cdot) = \pi_1(\cdot)$ in
the A0 algorithm to achieve $E^*$.  Such optimal A0 algorithms comprise a class, indexed
by $q_{10} \in [\frac{\pi_0}{\pi_1}, 1]$, with $q_{00} = 0, q_{01} = 1$, and $g(\theta) =
\pi_1(\theta)$ for all $\theta$.

By including a within-model move, many optimal A1 algorithms exist with $g(\cdot) \ne
\pi_1(\cdot)$ that achieve $E^*$ or generate the Markov chain of
eq.~\ref{eq:P-pi<1/2-kernel}.  Note that $P^*$ is achieved when conditions (a) and (b) of
eq.~\ref{eq:pi<1/2-conditions} are satisfied and all $0\to 1$ proposals are accepted. When
$q_{10} > \frac{\pi_0}{\pi_1}$, a range of proposal algorithms, $g(\cdot) \ne
\pi_1(\cdot)$ may satisfy condition (b).  The entrance distribution then matches the
proposal, and may differ from the target.  As a result, $\Pj$ may reach the maximum but
$E$ may be suboptimal.  Then including a within-model move to restore the target
distribution for $\theta$ in algorithm A1 will satisfy condition (c) and achieve both
$\Pj^*$ and $E^*$.  Such cases include the region $1 < a \le a^*$ in figure
\ref{fig:uniform-pjump-eff}\textbf{a}\&\textbf{c} for the uniform-distribution example,
and the round top of the table mountain of figure
\ref{fig:normal-pjump-eff}\textbf{a}\&\textbf{c} in the normal-distribution example.

\subsubsection{Case $\pi_0 \ge \frac{1}{2}$}

In this case, the optimal chain of eq.~\ref{eq:P-pi>1/2} is
\begin{equation} \label{eq:P-pi>1/2-discrete}
   P^* =
   \begin{bmatrix}
      \frac{2\pi_0-1}{\pi_0} & \frac{\pi_1}{\pi_0}\pi_1(\theta_1)\Delta_1 & \cdots & \frac{\pi_1}{\pi_0}\pi_1(\theta_K)\Delta_K \\
      1                       & 0                   & \cdots & 0 \\
      1                       & 0                   & \cdots & 0 \\
      \vdots                  & \vdots              & \ddots & \vdots \\
      1                       & 0                   & \cdots & 0
   \end{bmatrix},
\end{equation}
which achieves $\Pj^* = 2(1-\pi_0)$ and $E^* = 1/(2\pi_0 - 1)$. When $K \to \infty$, the
optimal chain on the continuous space has the kernel
\begin{equation} \label{eq:P-pi>1/2-kernel}
\begin{aligned}
   &P(0 \to 0)                    = 1 - \tfrac{\pi_1}{\pi_0}, \\
   &P(0 \to 1,\d\theta)           = \tfrac{\pi_1}{\pi_0} \pi_1(\d\theta), \\
   &P(1,\d\theta \to 0)           = 1, \\ 
   &P(1,\d\theta \to 1,\d\theta') = 0. 
\end{aligned}
\end{equation}

By considering the $1\to 0$ move, we have the following conditions for the optimal MCMC
algorithm
\begin{equation} \label{eq:pi>1/2-conditions}
   \begin{aligned}
      \text{(a) } &{} q_{10} = 1 \text{ (attempt model jump if in 1)};  \\
      \text{(b) } &{} A_{10}(\theta) = \tfrac{q_{01}\cdot g(\theta)}{q_{10}} \cdot \tfrac{\pi_0}{\pi_1 \pi_1(\theta)} 
      \ge 1, \ \ \text{for all } \theta .
   \end{aligned}
\end{equation}
A simple algorithm is to always attempt a model jump ($q_{01} = q_{10} = 1$) and
to propose $\theta$ from its target $\pi_1(\theta)$ in the $0\to 1$ move; in this
algorithm some of the $0\to1$ proposals are rejected.

By considering the $0\to 1$ move, the conditions of eq.~\ref{eq:pi>1/2-conditions} imply
\begin{equation} \label{eq:pi>1/2-conditions-2}
   \begin{aligned}
      (\text{a$'$) } &{} \tfrac{\pi_1}{\pi_0} \le q_{01} \le 1;  \\
      \text{(b$'$) } &{} A_{01}(\theta) = \tfrac{q_{10}}{q_{01}\cdot g(\theta)} \cdot \tfrac{\pi_1 \pi_1(\theta)}{\pi_0} 
      \le 1, \ \ \text{for all } \theta .
   \end{aligned}
\end{equation}
The minimum bound $(\frac{\pi_1}{\pi_0})$ for $q_{01}$ (for the Markov chain to remain
optimal) is given when the proposal matches the target, with $g(\theta) = \pi_1(\theta)$
for all $\theta$.  In case the proposal and target do not match (which may happen if
$q_{01} > \frac{\pi_1}{\pi_0}$) but the maximum $\Pj^*$ is achieved, condition (b$'$)
means that $\theta$ values generated from $g(\cdot)$ are `thinned' to have the
distribution $\pi_1(\cdot)$.  In other words, generation of $\theta$ during the $0\to 1$
move may be viewed an instance of rejection sampling: $\theta$ is drawn from $g(\theta)$
but accepted with probability $\frac{c\pi_1(\theta)}{g(\theta)}$ so that the accepted
values have the distribution $\pi_1(\theta)$ (where $c$ is a constant with
$\frac{c\pi_1(\theta)}{g(\theta)} \le 1$ for all $\theta$).

Thus if $\pi_0 \ge \frac{1}{2}$, a range of proposals $g(\cdot)$ in the A0 algorithm may
achieve $\Pj^*$ and $E^*$ at the same time, and maximum $\Pj$ is a necessary and
sufficient condition for maximum $E$.  The entrance distribution matches the target, with
$g(\cdot) \ne f_{01}^\text{in}(\cdot) = \pi_1(\cdot)$, and including a within-model move
in the A1 algorithm does not improve the efficiency any further (as the A0 algorithm is
already optimal).   See figure \ref{fig:uniform-pjump-eff}\textbf{g}--\textbf{i}, $\pi_0 =
0.7, a^* < a < 1$ in the uniform example, and the round top of the table mountain in
figure \ref{fig:normal-pjump-eff}\textbf{g}--\textbf{i} for the normal-distribution
example.  If algorithm A0 is not already optimal, including within-model moves in A1
improves the efficiency of the chain but efficiency will not reach $E^*$.

\section{SPR proposals in phylogenetics}

For $n$ species, there are $K=\frac{(2n-5)!}{2^{n-3} (n-3)!}$ possible un-rooted binary
trees, each with $n$ external branches and $(n-3)$ internal branches.  The cross-tree MCMC
algorithm uses tree-perturbation algorithms such as NNI and SPR (fig.~\ref{fig:nni-spr})
to move from one tree to another, generating a sample from the posterior distribution of
the tree topology with branch lengths.  Denote the state of the chain as $\omega = (k,
\theta_k)$ with $k \in \{1,2,\dots,K\}$ to be the tree index, and
$\theta_k=(t_1,t_2,\dots,t_{2n-3})$ to be the vector of branch lengths in tree $k$. If
there are parameters in the substitution model, they will be included in $\theta_k$ as
well.  We focus on the efficiency for estimating the posterior probability of the MAP tree
and the cross-tree jump probability $\Pj$.  We consider both within-tree moves that modify
branch lengths and cross-tree moves that use different strategies to propose alternative
tree topologies $(q_{kk'})$ and branch lengths in the new tree, $q(\theta_{k'} | k,
\theta_k, k')$.  

We tested the different algorithms using two datasets, one of $\psi\eta$-globin
pseudogenes from six primate species (fig.~\ref{fig:trees}\textbf{a}), and another of
mitochondrial (mt) genes from 29 mammalian species (fig.~\ref{fig:trees}\textbf{b}). For
the $\psi\eta$ dataset, there are $105$ possible un-rooted trees for six species.  The top
three trees differ from each other by an NNI move and in total have the posterior $\sim
100\%$.  For the mt dataset, the top two trees have posteriors 0.574 and 0.107, which
differ by an NNI move. We focus on the efficiency for estimating the posterior probability
of the MAP tree.  The results are summarized in tables
\ref{tab:psieta-primates}\&\ref{tab:mt-mammals}.

The baseline algorithm includes an SPR move
(fig.~\ref{fig:nni-spr}\textbf{b}\&\textbf{b'}) to update the tree and a within-tree
move to update branch lengths, which involves sampling a branch at random and change it
via a multiplier. We consider five variations to the within-tree move for updating
branch lengths, including A1 with one $(2n-3)$-D move to change all branch lengths; A2-1
with a sequence of 1-D moves to change the branch lengths; A2-2 with a sequence of 1-D
moves with an update of the tree length (sum of all branch lengths); A3-1 with a
sequence of 1-D Bactrian-Laplace moves to change branch lengths \citep{Yang2013}; and
A3-2 with a sequence of 1-D moves and a tree-length update using the Bactrian-Laplace
proposal.  See SI Extended Methods for details.  While the different within-tree
proposals are expected to affect the mixing of branch lengths, they showed little impact
on the mixing efficiency of the cross-tree algorithm (tables
\ref{tab:psieta-primates}\&\ref{tab:mt-mammals}).

We explored different strategies to propose alternative tree topologies and to generate
branch lengths for the new tree, including B1 (NNI$_0$) NNI with direct transfer of
branch lengths; B2 (NNI$_1$) NNI with a modification of the focal branch length; B3
(NNI$_5$) NNI with all five branch lengths around the focal branch modified using
multipliers; B4 (NNI$_\text{bw}$) NNI with branch weights; B5 (NNI$_\text{bw\&pw}$) NNI
with branch weights and parsimony weights; B6 (NNI$_\text{local}$) the local move of
\citep{Larget1999}; B7 (SPR$_{\text{bw}}$) SPR with branch weights; B8
(SPR$_\text{bw\&pw}$), SPR with both branch weights and parsimony weights.  See SI
Extended Methods for details.

Comparison of NNI-based moves suggests that it is better not to change the branch lengths
when moving between trees, but the differences are small.  All of B1 (NNI$_0$), B2
(NNI$_1$), and B3 (NNI$_5$) use direct transfer of the branch lengths around the focal
branch, and performed much better than B6 (NNI$_\text{local}$), which merges and splits
branches to form new branch lengths.  For example, B1 (NNI$_0$) is more efficient than B6
(NNI$_\text{local}$) by a factor of 20 ($=0.1778/0.0089$) for the $\psi\eta$ dataset, and
by a factor of 16.6 ($=0.0053/0.00032$) for the mt dataset.  It was previously noted that
direct transfer produces branch lengths closer to the maximum likelihood estimate (MLE)
and is thus preferable to merge-and-split approach (\citealp[][p.283-4]{Yang2014}; see
also fig.~\ref{fig:trees-psieta-ml-mp}).  NNI variants B1-B5 also performed much better
than SPR (the baseline and B7 SPR$_\text{bw}$).  For example, B1 (NNI$_0$) is more
efficient than baseline by a factor of 21 ($=0.1778/0.0086$) for the $\psi\eta$ dataset,
and by a factor of 20 ($=0.0053/0.00027$) for the mt dataset.  Those differences are due
to two factors.  First, as discussed above, NNI variants B1-B5 use direct transfer while
SPR uses merge-and-split to generate branch lengths for the new tree.  Second, NNI is a
`smaller' move than SPR, so that a higher proportion of proposed trees are
high-probability trees, leading to a higher model jump rate.

Using branch weights to preferentially change the tree around short internal branches
improves the cross-tree mixing efficiency for both NNI and SPR.  Efficiency for B4
(NNI$_\text{bw}$) is 2.2 (=0.3969/0.1778) times as high as for B1 (NNI$_0$) in the
$\psi\eta$ dataset, although the effect is much smaller in the mt dataset ($0.0057/0.0053
= 1.08$).  Similarly B7 (SPR$_\text{bw}$) is more efficient than B0 (SPR baseline) by a
factor of 2.0 (=0.0172/0.0086) in the $\psi\eta$ dataset, with a minor effect in the mt
dataset ($0.00028/0.00027 = 1.04$).  The differences between datasets may be because there
are large differences in internal branch lengths in the $\psi\eta$ tree
(Fig.~\ref{fig:trees}\textbf{a}) while branch lengths in the mt dataset are much more
homogeneous (Fig.~\ref{fig:trees}\textbf{b}).  Note that our branch weight
(eq.~\ref{eq:weight-branch}) is arbitrary and may suit one dataset better than another.

Using parsimony scores (eq.~\ref{eq:weight-parsimony}) to sample target branches improved
mixing efficiency.  In the NNI move, B5 NNI$_\text{bw\&pw}$ was more efficient than B4
NNI$_\text{bw}$ by a factor of 2.1 for the $\psi\eta$ dataset but no effect in the mt
dataset.  For the SPR move, B8 SPR$_\text{bw\&pw}$ was more efficient than B7
SPR$_\text{bw}$ by a factor of 3.5 ($= 0.0608/0.0172$) in the $\psi\eta$ dataset and a
factor of 1.7 ($= 0.00047/0.00028$) in the mt dataset.

Overall, tests using the two datasets confirm our theoretical analysis.  The greater
efficiency of direct transfer of branch lengths in the NNI move than the merge-split
approach confirms the importance of proposing model parameters (branch lengths) near the
mode of the within-model posterior for the new model.  Both the branch weights for
internal branches and the parsimony weights for the target branches alter the model
proposal probability ($q_{kk'}$), favoring trees that are close to the current tree and
are likely to have higher posteriors.  Those strategies are expected to improve the mixing
efficiency of cross-model algorithms from our theoretical analysis and are found to have a
major impact on the cross-tree mixing efficiency in both datasets.

We used B4 (NNI$_\text{bw}$) as the cross-tree move to study the optimal division of labor
between within-tree and cross-tree moves.  For the within-tree move we used Baseline,
which changes one branch length chosen at random.  For the $\psi\eta$-globin dataset, the
optimal cross-tree proposal probability was 0.7--0.8, while for the mt dataset, it was
$\sim$0.5 (Fig.~\ref{fig:pjump-eff-prop}\textbf{a}\&\textbf{b}).

\subsection{Division of computational efforts between within-tree and cross-tree moves}

\begin{figure*} [t]
   \centering
   \includegraphics[scale=1.11]{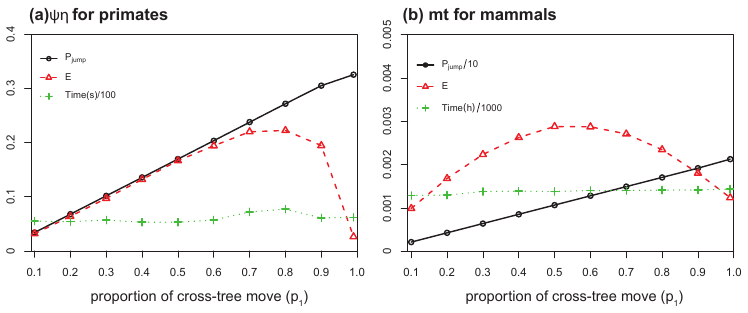}
   
   \caption{The cross-tree jump probability $\Pj$, the efficiency $E$ and running time
      plotted against the cross-tree proposal probability.  The cross-tree move is B4
      (NNI$_\mathrm{bw}$) while the within-tree move changes one randomly sampled branch
      length. %
   } \label{fig:pjump-eff-prop}
\end{figure*}

Instead of having a within-tree move and a cross-tree move in each MCMC iteration, here we
sample the cross-tree move with probability $p_1$ (and the within-tree move with $p_0 =
1-p_1$).  The optimal $p_1$ is around 0.75 for the $\psi\eta$ dataset and 0.55 for the mt
dataset (fig.~\ref{fig:pjump-eff-prop}).

\section{Discussion}

\subsection{Differences between within-model and trans-model MCMC algorithms}

There are a number of differences between within-model and trans-model algorithms
\citep{Yang2014, Nascimento2017}.  First, for a within-model move, one can make the step
size (e.g., the window size of the sliding-window proposal) small enough so that the
acceptance rate is $\sim 100\%$.  However, for trans-model moves, such a concept of a tiny
step does not exist, and furthermore the acceptance rate is constrained by the posterior
model probabilities (eq.~\ref{eq:pjump-bound}).  If the MAP model has the posterior 99\%,
$\Pj$ cannot exceed $2(1-0.99) = 2\%$, as otherwise the chain would not be visiting the
MAP model often enough to achieve the correct posterior. Second, for a within-model move,
intermediate acceptance rates, at 30-40\%, are optimal \citep{Gelman1996}.  For trans-model
moves, a mobile chain (with a high $\Pj$) is in general more efficient than a lazy chain
(with a low $\Pj$), and we should strive to achieve high acceptance rates, even though our
theory identified situations where maximum $\Pj$ is neither necessary nor sufficient for
maximum mixing efficiency.  While an acceptance rate of $\sim 0$ for a within-model
algorithm almost invariably indicates a mixing problem (e.g., the window size may be too
large), it may not necessarily mean a mixing problem for cross-model moves as it may be
due to extreme posterior probabilities. Whereas MCMC theory often emphasizes the general
applicability of the algorithms to complex parameter spaces, the differences between
within-model and trans-model algorithms should be accounted for when our aim is to develop
efficient trans-model algorithms.

Some proposals used in current phylogenetic programs are mixtures of within-tree moves and
cross-tree moves (e.g., B6 NNI$_\text{local}$), making it awkward to adjust the step
length to improve performance.  For example, Lakner et al.~\citep{Lakner2008} found that
the overall acceptance rate was not a good indicator of the efficiency of the algorithm,
whereas the acceptance rate of topology changes was.  This is apparently because the
acceptance rate in \citep{Lakner2008} averaged over within-tree moves (for which
intermediate acceptance rates are optimal) and cross-tree moves (for which high acceptance
rates are desirable), so that the overall acceptance rate is not a very useful indicator
of performance.

\subsection{Measures of mixing efficiency of MCMC algorithms in phylogenetics}

Besides the measure of efficiency based on the variance ratio (eq.~\ref{eq:eff}), an
alternative measure, used by \citep{Lakner2008} and \citep{Hohna2008}, is based on
probabilities for splits.  An extremely long `reference chain' is run to collect the
splits and their `true' probabilities.  One then runs the test MCMC algorithm for a
fixed number of iterations $(N)$, and calculate the distance between the `true split
probabilities' ($p_i$ for split $i$) and the estimates from the test chain ($\hat p_i$).
For instance, the distance may be defined as the maximum difference in split
probabilities,
\begin{equation}
  \delta = \max_i \bigl| \hat p_i - p_i \bigr|.
\end{equation}

This criterion is not very different from the variance ratio of eq.~\ref{eq:eff}.  Define
$f(k) = 1$ if tree $k$ contains a specific split and 0 otherwise.  Then $\E_\pi(f)$ will
be the posterior probability for the split, and the variance or standard deviation can be
used to measure the efficiency of the algorithm for estimating the split probability. Thus
the standard measure of mixing efficiency should be adequate for use in phylogenetics
(cf.~\citep{Meyer2021}).

\subsection{Efficiency of cross-model and cross-tree algorithms}

The cross-model algorithms analyzed in this paper, involving two models $H_0$ and $H_1$,
with $H_0$ having no free parameters, are among the simplest.  The optimal Markov chain
for estimating $\pi_0$ has different characteristics depending on $\pi_0$. If $\pi_0 \le
\frac{1}{2}$, the optimal chain moves to $H_1$ immediately when it is currently in $H_0$.
Maximum $\Pj$ is achieved for a range of proposal density $g(\cdot)$, but algorithm A0
without within-model moves is optimal achieving maximum $E$ only if $g(\cdot) =
\pi_1(\cdot)$.  If the cross-model move achieves the maximum $\Pj$ but not the maximum
$E$, including within-model moves (in algorithm A1) to restore $\theta$ to its stationary
distribution in $H_1$ will achieve maximum $E$.  If $\pi_0 > \frac{1}{2}$, it is possible
for an A0 algorithm with $g(\cdot) \ne \pi_1(\cdot)$ to achieve the maximum $E$. In this
case maximum $\Pj$ implies maximum $E$.  If the cross-model move with $g(\cdot) \ne
\pi_1(\cdot)$ does not achieve maximum $E$, including a within-model move (in algorithm
A1) will not either.  In both cases ($\pi_0 < \frac{1}{2}$ and $\pi_0 \ge \frac{1}{2}$),
there exist multiple cross-model MCMC algorithms that lead to the same Markov chain
achieving maximum $E$.

General cases where both models have free parameters, or where more than two models with
parameters are under comparison, are yet to be studied.  For example, it remains to be
shown that for estimating a posterior model probability, the efficiency will not be higher
than that achieved by proposing parameters from their posterior for the new model.

Similarly the cross-tree algorithms considered in this study are among the simplest.
Nevertheless, our tests using two real datasets highlight the important of both components
of the cross-tree proposal: the tree-proposal probability $q_{kk'}$, and the parameter
(branch-length) proposal $q(\theta_{k'}|k,\theta_k,k')$.

A cross-tree algorithm typically operates in two steps, both of which affect the
tree-proposal probability $q_{kk'}$: (i) an internal (focal) branch is selected at which a
subtree is pruned off or a local change is made, and (ii) a target branch on the
tree-backbone is selected for reattachment.  Weights based on internal branch lengths or
split probabilities \citep{Meyer2021} may be used to preferentially change the uncertain
parts of the current tree.  Here internal branch lengths are simple to use as they are
part of the current state of the Markov chain.  In contrast, split probabilities depend on
all past states, and as a result, one cannot use the current estimates of split
probabilities in the proposal, and long pilot runs have to be used to estimate them before
MCMC sampling starts \citep{Meyer2021}.  Similarly, features of the sequence data or of the
current tree that may be correlated with the posterior probabilities of the tree can be
used to sample target branches, such as parsimony scores \citep{Yang2014, Zhang2020,
   Meyer2021}, average sequence distances between species, and conditional probability
vectors for internal nodes in the unchanged parts of the tree calculated during the
post-order tree traversal algorithm \cite[][pp.103--106]{Yang2014}.  All these are
functions of the data or of the current state, and their use incurs simple Hastings
ratios.

Our empirical tests also highlight the importance of parameter proposal
$q(\theta_{k'}|k,\theta_k,k')$ in the cross-tree algorithms.  For the NNI move, direct
transfer of branch lengths was far more efficient than the merge-split approach,
apparently because it generates branch lengths close to their MLEs.  Direct transfer does
not apply to the SPR move, but it may be useful to sample the regrafting point on the
target branch non-uniformly to be closer to the pruning point, with $\E(x'/r) \ll
\frac{1}{2}$ in figure \ref{fig:nni-spr}\textbf{b}$'$.

Our study does not address the problem of multiple local peaks in the posterior space,
which appear to be common in phylogenetics \citep{Whidden2015}.  Indeed the `geometry' of
the tree space depends on the cross-tree algorithm (e.g., NNI vs.\ SPR).  For both
datasets, there appears to exist only a single `mode' in the space of trees, as all trees
with non-negligible posteriors appear to be connected to the MAP tree via NNI moves
(Figs.~\ref{fig:trees}\textbf{a}\&\textbf{b}).  In such cases, NNI may have lower
computational cost and may mix more efficiently than SPR or TBR (for tree-bisection-and
reconnection), which propose much larger changes to the tree with many candidate trees,
most of which are expected to be poor.  However, if multiple local peaks exist, SPR and
TBR are less likely to get stuck in a local peak as they allow the chain to move to
distant trees that are not NNI neighbors.  In practice a mixture of small (NNI) and large
(SPR and TBR) moves may be desirable. Tempering \citep{Jennison2003} and population MCMC
\citep{Jasra2007} are strategies for improve mixing when there exist multiple modes in the
within-model posterior.  Their effectiveness in cross-model algorithms is yet to be
studied.

We note that the efficiency calculated for the two real datasets is in general much lower
than that for the uniform and normal examples.  MCMC tree search in phylogenetics is
well-known to be inefficient.  It is common to run an MCMC algorithm for $N = 10^9$
iterations, say, achieving an ESS of $< 100$, in which case $E < 10^{-7}$.  We suggest
that this is partly because the tree models involve many parameters (branch lengths etc.),
and partly because current cross-tree algorithms have room for improvement.  Our theory
may serve as a useful guide when we design heuristic cross-tree algorithms with high
mixing efficiency.

\section{Materials and methods}

We tested different algorithms described in the paper using two datasets.  The first
consists of the $\psi\eta$-globin pseudogenes from six primate species: human, chimpanzee,
gorilla, orangutan, Rhesus monkey, and spider monkey (fig.~\ref{fig:trees}\textbf{a}).
There are 6,166 sites in the alignment.  The second dataset consists of a segment of the
mitochondrial genome from 29 mammalian species, with 1,799 sites in the alignment
(fig.~\ref{fig:trees}\textbf{b}). %

\section{Supplementary Material and Data Availability}

Supplemental information is available at \texttt{\small
   https://doi.org/10.5061/dryad.xxxxxxx}.

\section{Funding}

This study has been supported by a Biotechnology and Biological Sciences Research
Council grant (BB/Y004132/1, BB/X007553/1, BB/X018571/1) to Z.Y., and by a Shenzhen
Training Project of Excellent Scientific \& Technological Talents grant
(RCYX20221008093033012), Natural Science Foundation of China (NSFC) grant (12101295),
and Guangdong Natural Science Foundation grant (2022A1515011767) to X.J.

\bibliographystyle{natbib}
\renewcommand{\bibfont}{\footnotesize}
\bibliography{mcmc-transmodel}

\begin{thebibliography}{}

\bibitem[Bouckaert {\em et~al.}(2014)Bouckaert, Heled, Kuhnert, Vaughan, Wu,
  Xie, Suchard, Rambaut, and Drummond]{Bouckaert2014}
Bouckaert, R., Heled, J., Kuhnert, D., Vaughan, T., Wu, C.~H., Xie, D.,
  Suchard, M.~A., Rambaut, A., and Drummond, A.~J. 2014.
\newblock {BEAST} 2: a software platform for {Bayesian} evolutionary analysis.
\newblock {\em PLoS Comput. Biol.}, {10}(4): e1003537.

\bibitem[Brooks {\em et~al.}(2003)Brooks, Guidici, and Roberts]{Brooks2003}
Brooks, S.~P., Guidici, P., and Roberts, G.~O. 2003.
\newblock Efficient construction of reversible jump {Markov chain Monte Carlo}
  proposal distributions.
\newblock {\em J. R. Stat. Soc. B\/}, {65}: 3--39.

\bibitem[Capp\'e {\em et~al.}(2003)Capp\'e, Robert, and Ryd\'en]{Cappe2003}
Capp\'e, O., Robert, C.~P., and Ryd\'en, T. 2003.
\newblock Reversible jump {MCMC} converging to birth-and-death {MCMC} and more
  general continuous time samplers.
\newblock {\em J. R. Stat. Soc. B\/}, {65}: 679--700.

\bibitem[Carlin and Chib(1995)Carlin and Chib]{Carlin1995}
Carlin, B.~P. and Chib, S. 1995.
\newblock Bayesian model choice via {Markov chain Monte Carlo}.
\newblock {\em J. R. Stat. Soc. B\/}, {57}: 473--484.

\bibitem[Chen {\em et~al.}(2014)Chen, Kuo, and Lewis]{Chen2014}
Chen, M.-H., Kuo, L., and Lewis, P. 2014.
\newblock {\em Bayesian Phylogenetics: Methods, Algorithms, and
  Applications\/}.
\newblock Chapman and Hall/CRC, London.

\bibitem[Dellaportas {\em et~al.}(2002)Dellaportas, Forster, and
  Ntzoufras]{Dellaportas2002}
Dellaportas, P., Forster, J.~J., and Ntzoufras, I. 2002.
\newblock On {Bayesian} model and variable selection using {MCMC}.
\newblock {\em Stat. Comput.}, {12}: 27--36.

\bibitem[Felsenstein(2004)Felsenstein]{Felsenstein2004}
Felsenstein, J. 2004.
\newblock {\em Inferring Phylogenies\/}.
\newblock Sinauer Associates, Sunderland, Massachusetts.

\bibitem[Flouri {\em et~al.}(2018)Flouri, Jiao, Rannala, and Yang]{Flouri2018}
Flouri, T., Jiao, X., Rannala, B., and Yang, Z. 2018.
\newblock Species tree inference with {BPP} using genomic sequences and the
  multispecies coalescent.
\newblock {\em Mol. Biol. Evol.}, {35}(10): 2585--2593.

\bibitem[Flouri {\em et~al.}(2020)Flouri, Jiao, Rannala, and Yang]{Flouri2020}
Flouri, T., Jiao, X., Rannala, B., and Yang, Z. 2020.
\newblock A {Bayesian} implementation of the multispecies coalescent model with
  introgression for phylogenomic analysis.
\newblock {\em Mol. Biol. Evol.}, {37}(4): 1211--1223.

\bibitem[Frigessi {\em et~al.}(1992)Frigessi, Hwang, and Younes]{Frigessi1992}
Frigessi, A., Hwang, C.-R., and Younes, L. 1992.
\newblock Optimal spectral structure of reversible stochastic matrices, {Monte
  Carlo} methods and the simulation of {Markov} random fields.
\newblock {\em Ann. Appl. Prob.}, {2}(3): 610--628.

\bibitem[Gelman {\em et~al.}(1996)Gelman, Roberts, and Gilks]{Gelman1996}
Gelman, A., Roberts, G., and Gilks, W. 1996.
\newblock Efficient {Metropolis} jumping rules.
\newblock In J.~Bernardo, J.~Berger, A.~Dawid, and A.~Smith, editors, {\em
  Bayesian Statistics 5\/}, volume~5, pages 599--607. Oxford University Press,
  Oxford.

\bibitem[Geyer(1992)Geyer]{Geyer1992}
Geyer, C.~J. 1992.
\newblock Practical {Markov} chain {Monte Carlo}.
\newblock {\em Stat. Sci.}, {7}: 473--483.

\bibitem[Girolami and Calderhead(2011)Girolami and Calderhead]{Girolami2011}
Girolami, M. and Calderhead, B. 2011.
\newblock Riemann manifold {Langevin} and {Hamiltonian Monte Carlo} methods.
\newblock {\em J. R. Stat. Soc. B\/}, {73}(2): 123--214.

\bibitem[Godsill(2001)Godsill]{Godsill2001}
Godsill, S. 2001.
\newblock On the relationship between {Markov chain Monte Carlo} methods for
  model uncertainty.
\newblock {\em J. Comput. Graph. Stat.}, {10}: 1--19.

\bibitem[Green(1995)Green]{Green1995}
Green, P.~J. 1995.
\newblock Reversible jump {Markov} chain {Monte Carlo} computation and
  {Bayesian} model determination.
\newblock {\em Biometrika\/}, {82}: 711--732.

\bibitem[Green and Han(1992)Green and Han]{Green1992}
Green, P.~J. and Han, X.~L. 1992.
\newblock Metropolis methods, {Gaussian} proposals and antithetic variables.
\newblock In P.~Barone, A.~Frigessi, and M.~Piccioni, editors, {\em Stochastic
  Models, Statistical Methods and Algorithms in Image Analysis\/}, pages
  142--164. Springer, New York.

\bibitem[Grenander and Miller(1994)Grenander and Miller]{Grenander1994}
Grenander, U. and Miller, M.~I. 1994.
\newblock Representations of knowledge in complex systems.
\newblock {\em J. R. Stat. Soc. B\/}, {56}: 549--603.

\bibitem[Hastings(1970)Hastings]{Hastings1970}
Hastings, W.~K. 1970.
\newblock {M}onte {C}arlo sampling methods using {M}arkov chains and their
  applications.
\newblock {\em Biometrika\/}, {57}: 97--109.

\bibitem[Hohna and Drummond(2012)Hohna and Drummond]{Hohna2012}
Hohna, S. and Drummond, A.~J. 2012.
\newblock Guided tree topology proposals for {Bayesian} phylogenetic inference.
\newblock {\em Syst. Biol.}, {61}: 1--11.

\bibitem[Hohna {\em et~al.}(2008)Hohna, Defoin-Platel, and Drummond]{Hohna2008}
Hohna, S., Defoin-Platel, M., and Drummond, A.~J. 2008.
\newblock Clock-constrained tree proposal operators in {Bayesian} phylogenetic
  inference.
\newblock In {\em 8th IEEE International Conference on BioInformatics and
  BioEngineering\/}, page 10.1109/BIBE.2008.4696663, Athens (Greece). IEEE.

\bibitem[Jasra {\em et~al.}(2007)Jasra, Stephens, and Holmes]{Jasra2007}
Jasra, A., Stephens, D.~A., and Holmes, C.~C. 2007.
\newblock Population-based reversible jump {Markov chain Monte Carlo}.
\newblock {\em Biometrika\/}, {94}: 787--807.

\bibitem[Jennison {\em et~al.}(2003)Jennison, Hurn, and
  Al-Awadhi]{Jennison2003}
Jennison, C., Hurn, M.~A., and Al-Awadhi, F. 2003.
\newblock Discussion of a paper by {Brooks} et al.
\newblock {\em J. R. Stat. Soc. B\/}, {65}: 44--45.

\bibitem[Jukes and Cantor(1969)Jukes and Cantor]{Jukes1969}
Jukes, T. and Cantor, C. 1969.
\newblock Evolution of protein molecules.
\newblock In {\em {Munro, H.N.}, ed. Mammalian Protein Metabolism\/}, pages
  21--123. Academic Press, New York.

\bibitem[Kemeny and Snell(1960)Kemeny and Snell]{Kemeny1960}
Kemeny, J. and Snell, J. 1960.
\newblock {\em Finite Markov Chains\/}.
\newblock Princeton, Van Nostrand.

\bibitem[Lakner {\em et~al.}(2008)Lakner, van~der Mark, Huelsenbeck, Larget,
  and Ronquist]{Lakner2008}
Lakner, C., van~der Mark, P., Huelsenbeck, J.~P., Larget, B., and Ronquist, F.
  2008.
\newblock Efficiency of {Markov chain Monte Carlo} tree proposals in {Bayesian}
  phylogenetics.
\newblock {\em Syst. Biol.}, {57}: 86--103.

\bibitem[Larget and Simon(1999)Larget and Simon]{Larget1999}
Larget, B. and Simon, D. 1999.
\newblock Markov chain {Monte Carlo} algorithms for the {Bayesian} analysis of
  phylogenetic trees.
\newblock {\em Mol. Biol. Evol.}, {16}: 750--759.

\bibitem[Lartillot {\em et~al.}(2009)Lartillot, Lepage, and
  Blanquart]{Lartillot2009}
Lartillot, N., Lepage, T., and Blanquart, S. 2009.
\newblock {PhyloBayes} 3: a {Bayesian} software package for phylogenetic
  reconstruction and molecular dating.
\newblock {\em Bioinformatics\/}, {25}(17): 2286--2288.

\bibitem[Li {\em et~al.}(2000)Li, Pearl, and Doss]{Li2000}
Li, S., Pearl, D., and Doss, H. 2000.
\newblock Phylogenetic tree reconstruction using {Markov chain Monte Carlo}.
\newblock {\em J. Amer. Statist. Assoc.}, {95}: 493--508.

\bibitem[Lindley(1957)Lindley]{Lindley1957}
Lindley, D. 1957.
\newblock A statistical paradox.
\newblock {\em Biometrika\/}, {44}: 187--192.

\bibitem[Mau and Newton(1997)Mau and Newton]{Mau1997}
Mau, B. and Newton, M. 1997.
\newblock Phylogenetic inference for binary data on dendrograms using {Markov
  chain Monte Carlo}.
\newblock {\em J. Computat. Graph. Stat.}, {6}: 122--131.

\bibitem[Metropolis {\em et~al.}(1953)Metropolis, Rosenbluth, Rosenbluth,
  Teller, and Teller]{Metropolis1953}
Metropolis, N., Rosenbluth, A.~W., Rosenbluth, M.~N., Teller, A.~H., and
  Teller, E. 1953.
\newblock Equation of state calculations by fast computing machines.
\newblock {\em J. Chem. Phys.}, {21}: 1087--1092.

\bibitem[Meyer(2021)Meyer]{Meyer2021}
Meyer, X. 2021.
\newblock Adaptive tree proposals for {Bayesian} phylogenetic inference.
\newblock {\em Syst. Biol.}, {70}(5): 1015--1032.

\bibitem[Miyamoto {\em et~al.}(1987)Miyamoto, Slighton, and
  Goodman]{Miyamoto1987}
Miyamoto, M.~M., Slighton, J.~L., and Goodman, M. 1987.
\newblock Phylogenetic relations of humans and african apes from {DNA}
  sequences in the psi-eta-globin region.
\newblock {\em Science\/}, {238}: 369--373.

\bibitem[Nascimento {\em et~al.}(2017)Nascimento, dos Reis, and
  Yang]{Nascimento2017}
Nascimento, F.~F., dos Reis, M., and Yang, Z. 2017.
\newblock A biologist's guide to {Bayesian} phylogenetic analysis.
\newblock {\em Nature Ecol. Evol.}, {1}: 1446--1454.

\bibitem[Neal(2011)Neal]{Neal2011}
Neal, R.~M. 2011.
\newblock {MCMC} using {Hamiltonian} dynamics.
\newblock In S.~Brooks, A.~Gelman, G.~Jones, and X.-L. Meng, editors, {\em
  Handbook of Markov Chain Monte Carlo\/}, pages 113--162. Chapman and
  Hall/CRC, London.

\bibitem[Nishimura {\em et~al.}(2020)Nishimura, Dunson, and Lu]{Nishimura2020}
Nishimura, A., Dunson, D., and Lu, J. 2020.
\newblock Discontinuous {H}amiltonian {M}onte {C}arlo for discrete parameters
  and discontinuous likelihoods.
\newblock {\em Biometrika\/}, {107}(2): 365--380.

\bibitem[Peskun(1973)Peskun]{Peskun1973}
Peskun, P. 1973.
\newblock Optimum {Monte-Carlo} sampling using {Markov} chains.
\newblock {\em Biometrika\/}, {60}(3): 607--612.

\bibitem[Rannala and Yang(1996)Rannala and Yang]{Rannala1996}
Rannala, B. and Yang, Z. 1996.
\newblock Probability distribution of molecular evolutionary trees: a new
  method of phylogenetic inference.
\newblock {\em J. Mol. Evol.}, {43}: 304--311.

\bibitem[Rannala and Yang(2017)Rannala and Yang]{Rannala2017}
Rannala, B. and Yang, Z. 2017.
\newblock Efficient {Bayesian} species tree inference under the multispecies
  coalescent.
\newblock {\em Syst. Biol.}, {66}: 823--842.

\bibitem[Roberts and Rosenthal(2008)Roberts and Rosenthal]{Roberts1998}
Roberts, G. and Rosenthal, J. 2008.
\newblock Optimal scaling of discrete approximations to {Langevin} diffusions.
\newblock {\em J. R. Statist. Soc. B.}, {60}(255--226).

\bibitem[Ronquist {\em et~al.}(2012)Ronquist, Teslenko, van~der Mark, Ayres,
  Darling, Hohna, Larget, Liu, Suchard, and Huelsenbeck]{Ronquist2012}
Ronquist, F., Teslenko, M., van~der Mark, P., Ayres, D.~L., Darling, A., Hohna,
  S., Larget, B., Liu, L., Suchard, M.~A., and Huelsenbeck, J.~P. 2012.
\newblock {MrBayes} 3.2: Efficient {Bayesian} phylogenetic inference and model
  choice across a large model space.
\newblock {\em Syst. Biol.}, {61}: 539--542.

\bibitem[Sokal(1989)Sokal]{Sokal1989}
Sokal, A. 1989.
\newblock {\em Monte Carlo Methods in Statistical Mechanics: Foundations and
  New Algorithms\/}.
\newblock Lecture Notes for the Cours de Troisieme Cycle de la Physique en
  Suisse Romande, Lausanne, Switzerland (June 1989).

\bibitem[Stephens(2000)Stephens]{Stephens2000}
Stephens, M. 2000.
\newblock Bayesian analysis of mixture models with an unknown number of
  components--an alternative to reversible jump methods.
\newblock {\em Ann. Stat.}, {28}: 40--74.

\bibitem[Swofford {\em et~al.}(1996)Swofford, Olsen, Waddell, and
  Hillis]{Swofford1996}
Swofford, D., Olsen, G., Waddell, P., and Hillis, D. 1996.
\newblock Phylogeny inference.
\newblock In D.~M. Hillis, C.~Moritz, and B.~K. Mable, editors, {\em Molecular
  Systematics\/}, pages 407--514. Sinauer Associates, Sunderland,
  Massachusetts, 2 edition.

\bibitem[Thawornwattana {\em et~al.}(2018)Thawornwattana, Dalquen, and
  Yang]{Thawornwattana2018BA}
Thawornwattana, Y., Dalquen, D., and Yang, Z. 2018.
\newblock Designing simple and efficient {Markov chain Monte Carlo} proposal
  kernels.
\newblock {\em Bayesian Analysis\/}, {13}(4): 1033--1059.

\bibitem[Thomson and Brown(2022)Thomson and Brown]{Thomson2022}
Thomson, R.~C. and Brown, J.~M. 2022.
\newblock On the need for new measures of phylogenomic support.
\newblock {\em Syst. Biol.}, {71}: 917--920.

\bibitem[Whidden and Matsen(2015)Whidden and Matsen]{Whidden2015}
Whidden, C. and Matsen, F. A.~t. 2015.
\newblock Quantifying {MCMC} exploration of phylogenetic tree space.
\newblock {\em Syst. Biol.}, {64}(3): 472--491.

\bibitem[Yang(2014)Yang]{Yang2014}
Yang, Z. 2014.
\newblock {\em Molecular Evolution: A Statistical Approach\/}.
\newblock Oxford University Press, Oxford, UK.

\bibitem[Yang and Rannala(1997)Yang and Rannala]{Yang1997}
Yang, Z. and Rannala, B. 1997.
\newblock Bayesian phylogenetic inference using {DNA} sequences: a {Markov
  chain Monte Carlo} method.
\newblock {\em Mol. Biol. Evol.}, {14}: 717--724.

\bibitem[Yang and Rodriguez(2013)Yang and Rodriguez]{Yang2013}
Yang, Z. and Rodriguez, C.~E. 2013.
\newblock Searching for efficient {Markov chain Monte Carlo} proposal kernels.
\newblock {\em Proc. Natl. Acad. Sci. U.S.A.}, {110}(48): 19307--19312.

\bibitem[Yang and Yoder(2003)Yang and Yoder]{Yang2003}
Yang, Z. and Yoder, A. 2003.
\newblock Comparison of likelihood and {Bayesian} methods for estimating
  divergence times using multiple gene loci and calibration points, with
  application to a radiation of cute-looking mouse lemur species.
\newblock {\em Syst. Biol.}, {52}: 705--716.

\bibitem[Yang and Zhu(2018)Yang and Zhu]{Yang2018}
Yang, Z. and Zhu, T. 2018.
\newblock Bayesian selection of misspecified models is overconfident and may
  cause spurious posterior probabilities for phylogenetic trees.
\newblock {\em Proc. Natl. Acad. Sci. U.S.A.}, {115}(8): 1854--1859.

\bibitem[Zanella(2020)Zanella]{Zanella2020}
Zanella, G. 2020.
\newblock Informed proposals for local {MCMC} in discrete spaces.
\newblock {\em J. Am. Statist. Ass.}, {115}(530): 852--865.

\bibitem[Zhang {\em et~al.}(2020)Zhang, Huelsenbeck, and Ronquist]{Zhang2020}
Zhang, C., Huelsenbeck, J.~P., and Ronquist, F. 2020.
\newblock Using parsimony-guided tree proposals to accelerate convergence in
  {Bayesian} phylogenetic inference.
\newblock {\em Syst. Biol.}, {69}(5): 1016--1032.

\end{thebibliography}


\begin{figure*} [t]
   \centering
   \includegraphics[scale=0.70]{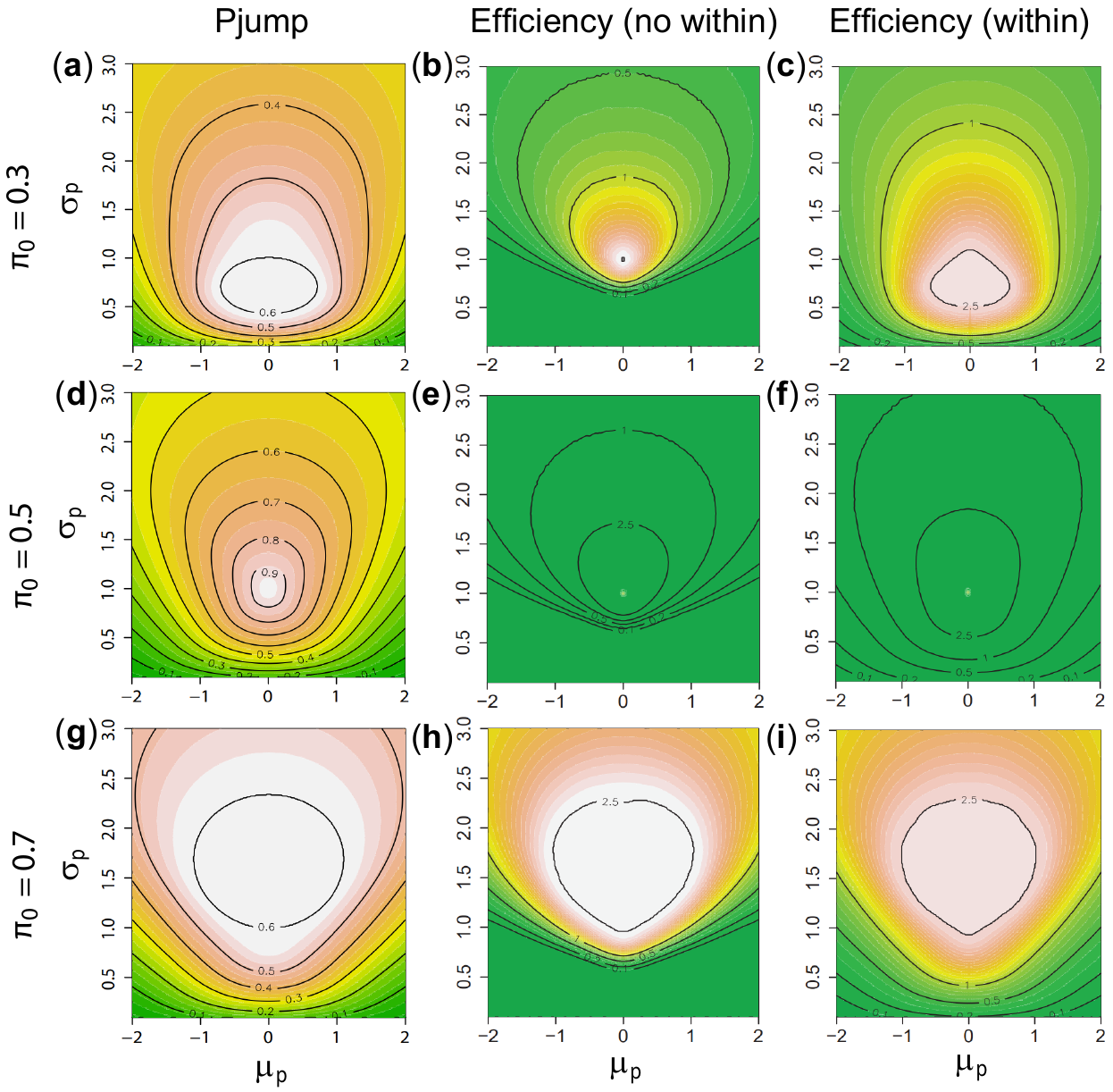}
   
   \caption{Contour plots of $\Pj$ and $E$ as functions of $(\mu_p, \sigma_p)$ in the
      proposal density $\mu \sim \N(\mu_p, \sigma_p^2)$ for the Normal-Distribution example.
      The three rows correspond to $\pi_0=0.3, 0.5, 0.7$.  $\Pj$ is calculated analytically
      while $E$ is by running the MCMC algorithm over $10^8$ iterations. \\ %
      (\textbf{a}) $\pi = 0.3$, maximum $\Pj^* = 0.6$ is achieved when $\mu_p$ and $\sigma_p$
      are in the region given by eq.~\ref{eq:N-round-table-pi<0.5}, represented by the flat
      top of the table mountain.  The region represents a class of MCMC algorithms.  For each
      of them, every move from $H_0$ to $H_1$ is accepted, with $A_{01}\ge 1$. \\ %
      (\textbf{b}) Algorithm A0 without within-model move achieves maximum efficiency $E^* =
      2.5$ only at one point $(\mu_p, \sigma_p) = (0,1)$. Elsewhere in the region, $E<E^*$ as
      $g(\cdot) = f_{01}^\text{in}(\cdot) \ne \pi_1(\cdot)$.  (\textbf{c}) Algorithm A1
      including a within-model move achieves $E^* = 2.5$ for the class of algorithms
      represented by the flat top defined by eq.~\ref{eq:N-round-table-pi<0.5}. Note that the
      round flat tops in (\textbf{a}) and (\textbf{c}) are the same, the apparent differences
      being due to numerical issues in calculation of $E$ and in kernel density smoothing. \\
      (\textbf{d}--\textbf{f}) When $\pi = 0.5$, the maximum $\Pj^* = 1$ and maximum $E^* =
      \infty$ are achieved at only one point $(\mu_p, \sigma_p) = (0,1)$. \\ %
      (\textbf{g}--\textbf{i}) $\pi_0 = 0.7$, maximum $\Pj^* = 0.6$ is achieved when $\mu_p$
      and $\sigma_p$ are in a region given by eq.~\ref{eq:N-round-table-pi>0.5}, represented
      by the flat top of the table mountain.  In this region, $A_{01} \le 1$ for all $\mu$ and
      as a result, algorithm A0 without the within-model move already achieves $E^*=2.5$, so
      that including a within-model move in A1 does not have an effect. %
   }	\label{fig:normal-pjump-eff}
\end{figure*}

\begin{figure*} [t]
   \centering
   \includegraphics[scale=0.60]{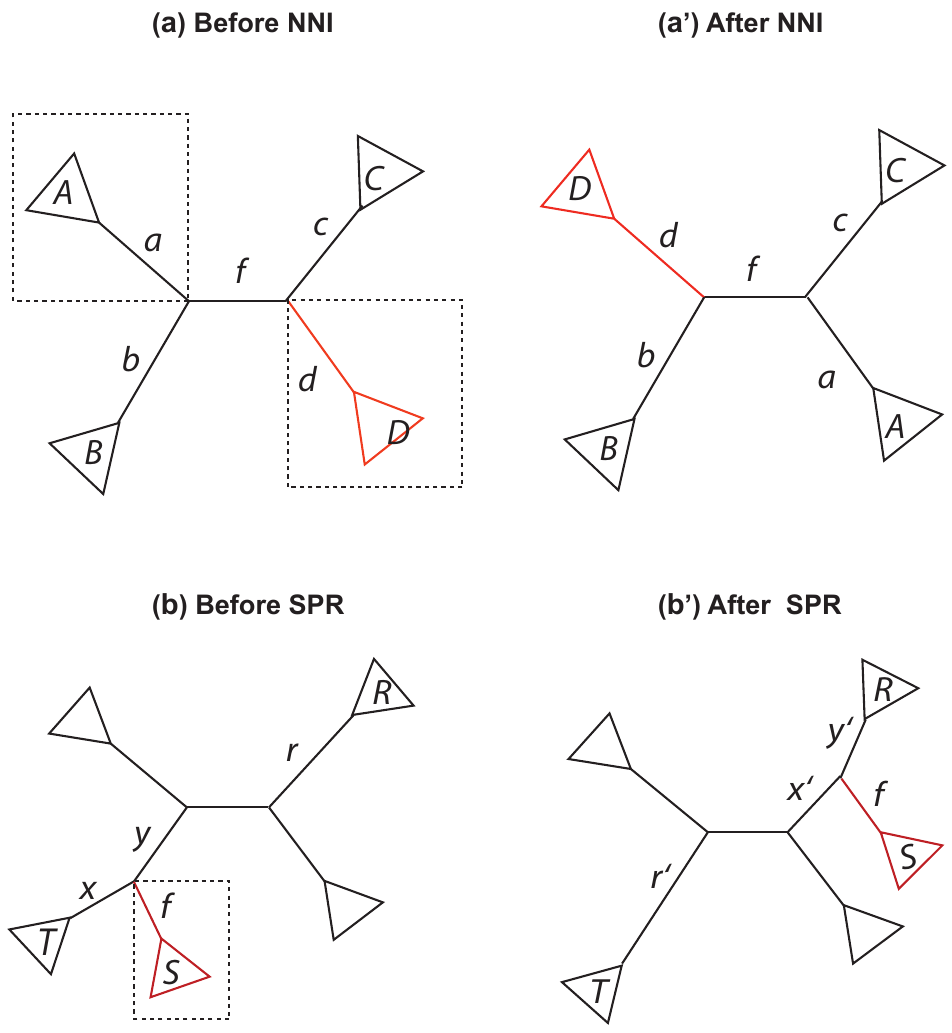}
   
   \caption{(\textbf{a}\&\textbf{a}$'$) The nearest-neighbor-interchange (NNI) move
      picks an internal branch ($f$) at random as the focal branch and swaps two
      randomly-chosen nodes from its two ends (e.g., $a$ and $d$).  Here the branch lengths
      are directly transferred from the current tree to the new tree. %
      (\textbf{b}\&\textbf{b'}) The subtree-pruning and regrafting (SPR) move prunes off a
      random branch $f$ and the connected subtree $S$ and re-grafts it to the backbone of
      the tree (e.g., onto branch $r$).  Here the move merges branches $x$ and $y$ into one
      branch ($r'$) and splits branch $r$ into two branches ($x', y'$).  %
   } \label{fig:nni-spr}
\end{figure*}

\begin{figure*} [t]
   \centering
   \includegraphics[scale=0.60]{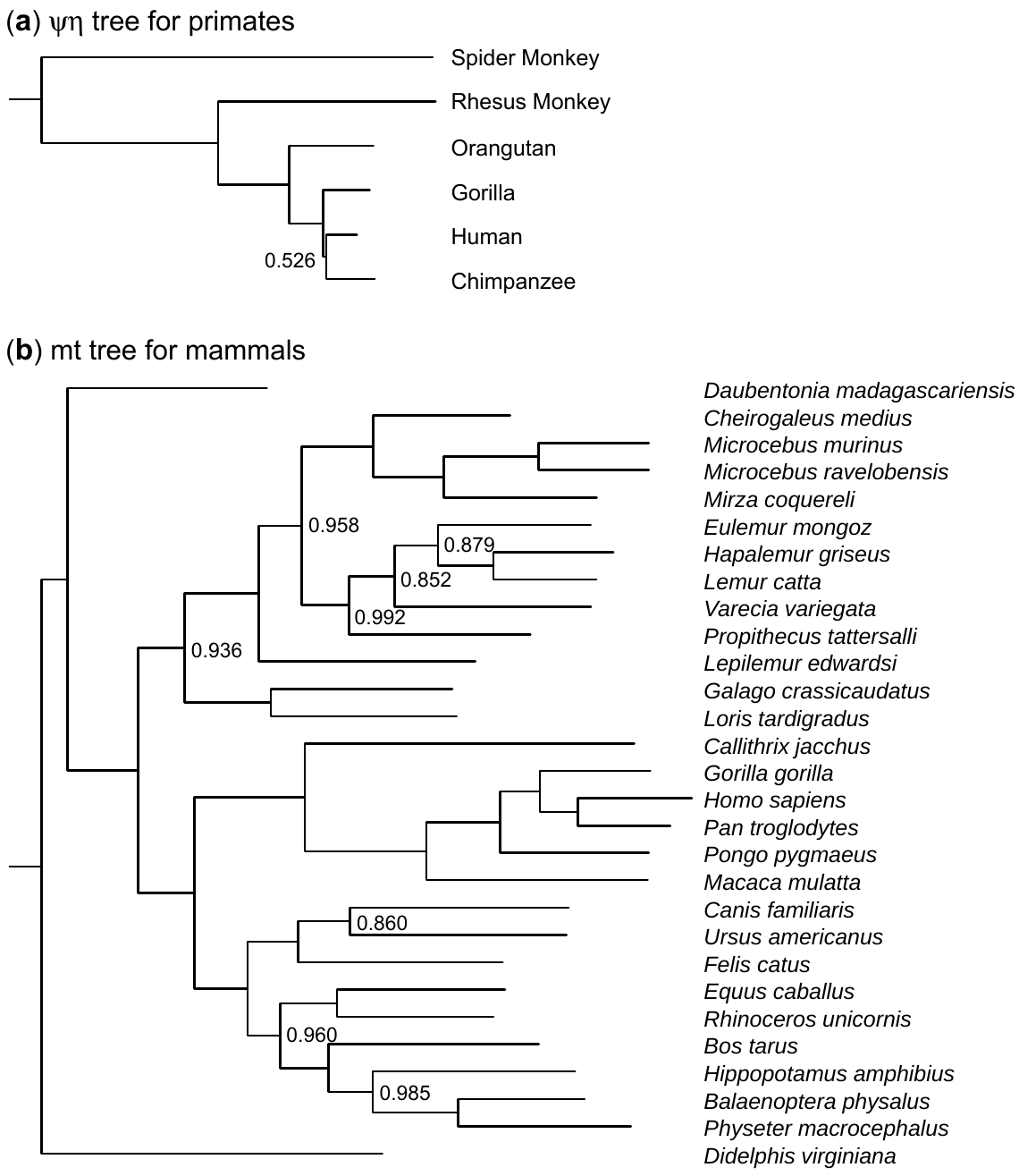}
   
   \caption{(\textbf{a}) The MAP tree for six primate species for the $\psi\eta$-globin
      pseudogenes. The branches are drawn to represent the posterior means of their lengths.
      The tree is unrooted and the root is placed on the spider monkey branch for clarity.
      The top three trees differ regarding the relationships among the human (H), chimpanzee
      (C) and gorilla (G), with the posterior 0.526 for ((HC)G), 0.438 for ((CG)H), and 0.035
      for ((HG)C).  The other two internal branches had 100\% posterior support. %
      (\textbf{b}) The MAP tree for 29 mammalian species for the mitochondrial dataset
      (M1510YY03).  The numbers next to internal nodes are posterior probabilities for the
      corresponding splits.  The second best tree differs from the MAP tree around the node
      highlighted and groups \textit{Eulemur mongoz} with \textit{Varecia variegata}.  The
      posterior probability for the whole tree is 0.58. %
   } \label{fig:trees}
\end{figure*}

\FloatBarrier

\begin{table*}[!htbp]
   \centering %
   \caption{Most efficient Markov chain for estimating a posterior model probability
      when the target distribution is $\pi = (\frac{4}{7}, \frac{2}{7}, \frac{1}{7})$ } %
   \label{tab:P3x3-example} %
   \begin{tabular}{ @{} lcccccccc @{} }
      \toprule
      Estimating & $f$ & $P^*$ & $\nu$ & $\Pj^*$ & $E^*$ & Eigenvalues \\
      \midrule
      
      $\pi_1 = \frac{4}{7}$ &
      $\begin{bmatrix}
         1 \\ 0 \\ 0 \\
      \end{bmatrix}
      \to
      \begin{bmatrix}
         \frac{3}{7} \\ -\frac{4}{7} \\ -\frac{4}{7} \\
      \end{bmatrix}$
      &
      \setlength{\arraycolsep}{2pt}
      $\begin{bmatrix}
         \frac{1}{4} & \frac{2}{4} & \frac{1}{4} \\
         1           & 0           & 0 \\
         1           & 0           & 0 \\
      \end{bmatrix} $ %
      &  $\frac{12}{343}$  & $\frac{6}{7}$ & $\frac{1}{2\pi_1-1} = 7$ & $(1,0,-\frac{3}{4})$ \\ \\
      
      $\pi_2 = \frac{2}{7}$ &
      $\begin{bmatrix}
         0 \\ 1 \\ 0 \\
      \end{bmatrix}
      \to
      \begin{bmatrix}
         -\frac{2}{7} \\ \frac{5}{7} \\ -\frac{2}{7} \\
      \end{bmatrix}$
      &
      \setlength{\arraycolsep}{2pt}
      $\begin{bmatrix}
         \frac{12}{25} & \frac{10}{25} & \frac{3}{25} \\
         \frac{20}{25} & 0             & \frac{5}{25} \\
         \frac{12}{25} & \frac{10}{25} & \frac{3}{25} \\
      \end{bmatrix}
      $
      &  $\frac{30}{343}$  & $\frac{124}{175}$ & $\frac{1}{1-2\pi_2} = \frac{7}{3}$ & $(1,0,-\frac{2}{5})$ \\ \\
      
      $\pi_3 = \frac{1}{7}$ &
      $\begin{bmatrix}
         0 \\ 0 \\ 1 \\
      \end{bmatrix}
      \to
      \begin{bmatrix}
         -\frac{1}{7} \\ -\frac{1}{7} \\ \frac{6}{7} \\
      \end{bmatrix}$
      &
      \setlength{\arraycolsep}{2pt}
      $\begin{bmatrix}
         \frac{10}{18} & \frac{5}{18} & \frac{3}{18} \\
         \frac{10}{18} & \frac{5}{18} & \frac{3}{18} \\
         \frac{12}{18} & \frac{6}{18} & 0 \\
      \end{bmatrix}
      $
      &  $\frac{30}{343}$  & $\frac{38}{63}$ & $\frac{1}{1-2\pi_3} = \frac{7}{5}$ & $(1,0,-\frac{1}{6})$ \\ %
      \bottomrule
   \end{tabular} 
   
   Note.--- The function $f$ for estimating $\pi_1, \pi_2, \pi_3$ is centered to have mean
   0, with $\sum_k \pi_k f_k = 0$.  Efficiency is defined as the variance ratio, $E =
   \nu_f/\nu$, with $\nu_f = \pi_k(1-\pi_k)$ for estimating $\pi_k$ (eq.\ref{eq:eff}). The
   eigenvalues are those of $P^*$. \\ %
\end{table*}

\begin{table*} [!htbp]
   \centering %
   \caption{Performance of different algorithms using the $\psi\eta$-globin dataset} %
   \small%
   \begin{tabular}{ l ccccccccc }
      \toprule
      & Baseline & A1 & A2-1 & A2-2 & A3-1 & A3-2 \\
      \midrule
      \multicolumn{3}{l}{Within-tree algorithms} \\
      $\Pj$ & 0.0102 & 0.0102 & 0.0102 & 0.0102 &  0.0102 & 0.0102\\
      $E$ & 0.0086 & 0.0083 & 0.0088 & 0.0089 & 0.0090 & 0.0089\\
      $\text{Pr(MAP tree)}$ & 0.5261 & 0.5242 & 0.5259 & 0.5260 & 0.5240 & 0.5270\\
      $\text{Pr(2nd Tree)}$ & 0.4373 & 0.4398 & 0.4381 & 0.4386 & 0.4397 & 0.4364\\
      Time (s) & 8.2 & 10.0 & 24.2 & 27.2 & 24.6 & 27.4 \\
      \\
      \multicolumn{3}{l}{Cross-tree algorithms} \\
      & Baseline            & B1 NNI$_0$ & B2 NNI$_1$ & B3 NNI$_5$ & B4 NNI$_\text{bw}$ 
      & B5 NNI$_\text{bw\&pw}$ & B6 NNI$_\text{local}$ & B7 SPR$_\text{bw}$ & B8 SPR$_\text{bw\&pw}$ \\
      $\Pj$                 & 0.0102 & 0.1689 & 0.1690 & 0.1687 & 0.3388 & 0.5569 & 0.0104 & 0.0212 & 0.0660\\
      $E$                   & 0.0086 & 0.1778 & 0.1765 & 0.1762 & 0.3969 & 0.8188 & 0.0089 & 0.0172 & 0.0608 \\
      $\text{Pr(MAP tree)}$ & 0.5261 & 0.5262 & 0.5265 & 0.5261 & 0.5262 & 0.5258 & 0.5240 & 0.5259 & 0.5249 \\
      $\text{Pr(2nd Tree)}$ & 0.4373 & 0.4376 & 0.4374 & 0.4376 & 0.4376 & 0.4378 & 0.4403 & 0.4373 & 0.4386 \\
      Time (s)              & 8.2    & 8.0    & 8.1    & 8.6    & 7.9    & 8.8    & 8.3    & 8.1    & 8.2 \\
      \bottomrule
   \end{tabular}   \label{tab:psieta-primates}
\end{table*}

\begin{table*} [!htbp]
   \centering %
   \caption{Performance of different algorithms using the mitochondrial dataset
      (M1510YY03) } %
   \small%
   \setlength{\tabcolsep}{3pt}
   \begin{tabular}{ lccccccccc }
      \toprule
      & Baseline & A1 & A2-1 & A2-2 & A3-1 & A3-2 \\
      \midrule
      \multicolumn{3}{l}{Within-tree algorithms} \\
      $P_\text{jump}$ & $6.3\times 10^{-5}$ & $6.3\times 10^{-5}$ & $6.4\times 10^{-5}$ & $6.4\times 10^{-5}$ & $6.4\times 10^{-5}$ & $6.4\times 10^{-5}$ \\
      $E$ & 0.00027 & 0.00027 & 0.00027 & 0.00027 & 0.00027 & 0.00027 \\
      $\text{Pr(MAP tree)}$ & 0.575 & 0.581 & 0.570 & 0.576 & 0.576 & 0.578 \\
      $\text{Pr(2nd Tree)}$ & 0.110 & 0.110 & 0.111 & 0.109 & 0.111 & 0.105 \\
      Time (h) & 2:20:54 & 3:51:35 & 52:48:17 & 54:41:55 & 52:39:59 & 54:34:01 \\ \\
      
      \multicolumn{3}{l}{Cross-tree algorithms} \\
      & Baseline & B1 NNI$_0$ & B2 NNI$_1$ & B3 NNI$_5$ & B4 NNI$_\text{bw}$ 
      & B5 NNI$_\text{bw\&pw}$ & B6 NNI$_\text{local}$ & B7 SPR$_\text{bw}$ & B8 SPR$_\text{bw\&pw}$ \\
      $P_\text{jump}$       & $6.3\times 10^{-5}$ & 0.018 & 0.018 & 0.018 & 0.021 & 0.038 & 0.00027 & $9.8\times 10^{-5}$  & 0.00147 \\
      $E$                   & 0.00027 & 0.0053  & 0.0052  & 0.0052  & 0.0057  & 0.0053 & 0.00032 & 0.00028 & 0.00047 \\
      $\text{Pr(MAP tree)}$ & 0.575   & 0.576   & 0.575   & 0.576   & 0.576   & 0.575   & 0.581   & 0.578   & 0.575 \\
      $\text{Pr(2nd Tree)}$ & 0.110   & 0.108   & 0.108   & 0.109   & 0.109   & 0.109   & 0.107   & 0.108   & 0.108 \\
      Time (h)              & 2:20:54 & 2:10:36 & 2:09:32 & 2:10:14 & 2:10:11 & 2:05:02 & 2:13:15 & 2:16:47 & 3:07:26 \\
      \bottomrule
   \end{tabular}
   \label{tab:mt-mammals}
\end{table*}

\clearpage
\section{Supplemental Information}

\textbf{\Large Mixing efficiency of trans-model Markov chain Monte Carlo algorithms with
	applications in Bayesian phylogenetics} \\%

\noindent%
\textbf{Xiyun Jiao, Tom\'{a}\v{s} Flouri, and Ziheng Yang} %


\begin{itemize}
	\item SI text 1. Proof of theorem 1
	\item SI text 2. Autocorrelation functions with maximum $\Pj$ but suboptimal
	efficiency in the case of $\pi_0 < \frac{1}{2}$ 
	\item SI text 3. Analysis of the Normal-Distribution example
	\item SI text 4. Super-efficient Markov chains in the discrete case
	\item Extended Methods
\end{itemize}

\renewcommand{\thefigure}{S\arabic{figure}}
\renewcommand{\thetable}{S\arabic{table}}
\renewcommand{\thepage}{S\arabic{page}}
\renewcommand{\theequation}{S\arabic{equation}}
\setcounter{figure}{0}
\setcounter{table}{0}
\setcounter{page}{1}

\section{SI text 1. Proof of theorem 1} 

Here we prove that $ \Pj \le 2 (1 - \max\{\pi_k\})$ (eq.~\ref{eq:pjump-bound}).  Without
loss of generality, we assume that the first model is the most probable so that
$\max\{\pi_k\} = \pi_1$.  If $\pi_1 \le \frac{1}{2}$, the inequality is obviously true. 
Note that in this case Markov chains achieving $\Pj=100\%$ are not unique.

Here we show that if $\pi_1 > \frac{1}{2}$, then $\Pj < 2(1-\pi_1)$.  First, consider
the case of no parameters in any of the models.  We have
\begin{equation} \label{eq:pjump-bound-1}
\Pj = \sum_{k=1}^K \pi_k (1-p_{kk}) = \pi_1 (1-p_{11}) + \sum_{k \neq 1} \pi_k (1-p_{kk}) .
\end{equation}
Because $\pi = (\pi_1,\dots,\pi_K)$ is the stationary distribution,
\begin{equation} \label{eq:pjump-bound-2}
\sum_{k=1}^K \pi_k p_{k1} = \pi_1 p_{11}+\sum_{k \neq 1} \pi_k p_{k1}=\pi_1,
\end{equation}
\begin{equation} \label{eq:pjump-bound-3}
\pi_1 (1-p_{11})=\sum_{k \neq 1} \pi_k p_{k1}\le\sum_{k \neq 1} \pi_k=1-\pi_1.
\end{equation}
Furthermore,
\begin{equation} \label{eq:pjump-bound-4}
\sum_{k \neq 1} \pi_k (1-p_{kk})\le \sum_{k \neq 1} \pi_k=1-\pi_1
\end{equation}
Thus $\Pj \le 2(1-\pi_1)$.

Second, if there are unknown parameters in the models, we replace $p_{kk'}$ with
$p_{kk'}^*$ of eq.~\ref{eq:pkk}.  Then
eqs.~\ref{eq:pjump-bound-1}--\ref{eq:pjump-bound-4} hold and inequality
\ref{eq:pjump-bound} follows.

The proof above does not use the detailed-balance condition and applies to both reversible
and non-reversible chains.  In the case of $\pi_1 > \frac{1}{2}$, equality in
eqs.~\ref{eq:pjump-bound-3} \& \ref{eq:pjump-bound-4} is achieved if and only if $p_{k1} =
1$ for all $k \neq 1$, and for reversible chains, this is equivalent to $p_{1k} =
\pi_k/\pi_1$ for all $k \neq 1$. In other words, the maximum $\Pj = 2(1 - \pi_1)$ (in the
case $\pi_1 > \frac{1}{2}$) is achieved by moving to model 1 immediately if the current
state is not 1.

\newpage
\section{SI text 2. Autocorrelation functions with maximum $\Pj$ but suboptimal
	efficiency in the case of $\pi_0 < \frac{1}{2}$ }

Consider two models, $H_0$ with no parameters and $H_1$ with parameters $\theta$, with the
target probabilities $\pi_0$ and $\pi_1$, and the distribution $\pi_1(\theta)$ in $H_1$.
If $\pi_0 \ge \frac{1}{2}$, there is a unique Markov chain that achieves the maximum
$\Pj^* = 2(1-\pi_0)$ and also the maximum efficiency for estimating $\pi_0$: $E^* =
1/(2\pi_0 - 1)$.  However, if $\pi_0 < \frac{1}{2}$, the Markov chain may achieve the
maximum $\Pj^* = 2\pi_0$, by moving to $H_1$ immediately when it is in $H_0$, but may not
achieve the maximum efficiency, $E^* = 1/(1 - 2\pi_0)$, for estimating $\pi_0$.  Here we
demonstrate that this occurs because the lag-$1$ autocorrelation in model index ($\rho_1$)
achieves the optimal (minimum) value ($\rho^*$) but the other autocorrelations ($\rho_k$,
$k\ge 2$) are not optimal.

We consider only a cross-model move in each MCMC iteration.  If the Markov chain is in
model 0, it moves to 1 immediately, with $A_{01} \ge 1$ for all $\theta$, whereas if the
chain is in 1, it moves to 0 with probability
\begin{equation}
A_{10} = g(\theta) \frac{\pi_0}{\pi_1 \pi_1(\theta)} \le 1, \quad \text{for all } \theta.
\label{eq:A01}
\end{equation}

Let $X_t$ be the model index at iteration $t$, taking 0 for model 0 and 1 for model 1. The
autocorrelation $\rho_k$ for estimating $\pi_0$ can be defined by using a function that
takes the value 1 for model 0 and 0 for model 1: $Y_t = 1$ (or 0) if $X_t = 0$ (or 1).  We
define two conditional probabilities: $Q_k = \mathbb{P}\{X_{t+k} = 0 | X_t = 0 \}$ and $R_k
= \mathbb{P}\{X_{t+1} = 1, \cdots, X_{t+k-1} = 1, X_{t+k} = 0 | X_t = 0 \}$.
Both $Q_k$ and $R_k$ are for a sequence of model indices with two 0s $k$-steps apart, but in
$R_k$ the intermediate steps are all 1s while in $Q_k$ they can be either 0 or 1.  Then
\begin{equation} \label{eq:rho}
\rho_k = \frac{\mathbb{E}\{Y_t Y_{t+k}\} - \mathbb{E}^2\{Y_t\}}{\mathbb{V}\{Y_t\}}
= \frac{\pi_0 Q_k - \pi_0^2}{\pi_0 \pi_1} = \frac{Q_k - \pi_0}{\pi_1}.
\end{equation}

$R_k$ is given by the geometric probability
\begin{equation} \label{eq:Rk}
\begin{aligned}
R_k &= \int g(\theta) (1 - A_{10})^{k-2} A_{10} \ \d \theta  \\
&= \frac{\pi_0}{\pi_1} \int \frac{g^2(\theta)}{\pi_1(\theta)}
\sum_{i=0}^{k-2} {k-2 \choose i}
\biggl( \frac{-\pi_0 g(\theta)}{\pi_1 \pi_1(\theta)} \biggr)^i  \d \theta  \\
&= \sum_{i=0}^{k-2} (-1)^i {k-2 \choose i} c_{i+2} \times \Bigl( \frac{\pi_0}{\pi_1} \Bigr) ^{i+1},
\end{aligned}
\end{equation}
where
\begin{equation} \label{eq:rho_k}
c_k = \int \frac{g(\theta)^k}{\pi_1(\theta)^{k-1}} \d \theta.
\end{equation}

For the uniform example, with $g(\theta)$ from beta($a,a$) and $1 \le a
\le a^* $ in the proposal, we have
\begin{equation}
c_k = \frac{B(a k - k + 1, a k - k + 1)}{B(a, a)^k}.
\end{equation}

For the Gaussian example, with $\frac{\pi_0}{\pi_1} \le \sigma_p \le 1$ and $|\mu_p| \le
\sqrt{2(1 - \sigma_p^2) \log \frac{\pi_1\sigma_p}{\pi_0}}$ in the proposal $g(\mu) =
\phi(\mu; \mu_p, \sigma_p^2)$, we have
\begin{equation}
c_k = \frac{ \exp\Bigl\{ \frac{k(k-1)\mu_p^2}{2[k-(k-1)\sigma_p^2]}
	\Bigr\} }
{\sigma_p^{k-1}\sqrt{k-(k-1)\sigma_p^2}}.
\end{equation}
If $\mu_p = 0$, this becomes $c_k = \frac{1}{\sigma_p^{k-1}\sqrt{k-(k-1)\sigma_p^2}}$.

Given $R_k$, we have
\begin{equation} \label{eq:Qk}
\begin{aligned}
&Q_1 = R_1 = 0,  \ \text{ for 00},    \\
&Q_2 = R_2,      \ \text{ for 010},   \\
&Q_3 = R_3,      \ \text{ for 0110},  \\
&Q_4 = R_4 + R_2^2,     \ \text{ for 01110, 01010},  \\
&Q_5 = R_5 + 2R_3 R_2,  \ \text{ for 011110, 011010, 010110},  \\
&Q_6 = R_6 + 2R_4 R_2 + R_3^2 + R_2^3, \\
&\cdots.
\end{aligned}
\end{equation}
Here the terms in $Q_k$ are given by a partition function that expresses $k$ as a sum of
integers $\ge 2$, and the coefficient for each term is the number of permutations for
those integers.  For example $k = 6$ can be partitioned in four ways, $6 = 4 + 2 = 3
+ 3 = 2 + 2 + 2$, so that $Q_6$ is a sum of four terms.  The second term ($2 R_4 R_2$)
corresponds to the two permutations of the runs 01110 and 010, generating the sequences
0111010 and 0101110.

Combining eqs.~\ref{eq:rho}, \ref{eq:Rk} \& \ref{eq:Qk}, we have
\begin{equation} \label{eq:rho_k2}
\begin{aligned}
&\rho_1 = -\frac{\pi_0}{\pi_1},  \\
&\rho_2 = -\frac{\pi_0}{\pi_1} + c_2 \pi^{(2)},  \\
&\rho_3 = \rho_2 -  c_3 \pi^{(3)}, \\
&\rho_4 = \rho_2 - (2c_3-c_2^2) \pi^{(3)} + c_4 \pi^{(4)}, \\
&\rho_5 = \rho_2 - (3c_3-2c_2^2) \pi^{(3)} + (3c_4 - 2c_2c_3) \pi^{(4)} - c_5 \pi^{(5)}, \\
&\rho_6 = \rho_2 - (4c_3-3c_2^2) \pi^{(3)} + (6c_4 - 6c_2c_3 +c_2^3) \pi^{(4)} \\
&\phantom{++} - (4c_5-2c_2c_4-c_3^2) \pi^{(5)} + c_6 \pi^{(6)}, \\
&\cdots,
\end{aligned}
\end{equation}
where $\pi^{(k)} = \pi_0^{k-1} / \pi_1^k$.

Note that if $g(\theta) = \pi_1(\theta)$, we have $c_k = 1$, and then $\rho_k^* = \bigl(
-\frac{\pi_0}{\pi_1} \bigr)^k$ for $k \ge 1$, and $E^* = \frac{1}{1 - 2\pi_0}$.

\newpage
\section{SI text 3. Analysis of the Normal-Distribution example} 

Suppose $H_0$ has no free parameters and $H_1$ has one parameter $\mu$, with the
posterior probabilities $\pi_0$ and $\pi_1 = 1 - \pi_0$ for the two models, and $\mu
\sim \N(0, 1)$ within $H_1$.  We use $g(\mu) = \phi(\mu; \mu_p, \sigma_p^2)$ as the
proposal density when the chain moves from 0 to 1, where $\phi(z; \mu, \sigma^2)$ is the
probability density function (PDF) for $\N(\mu, \sigma^2)$.  Algorithm A0 thus consists
of one cross-model move with $q_{01} = q_{10} = 1$ and $q(\mu|0,1) = \phi(\mu; \mu_p,
\sigma_p^2)$.  Then
\begin{align} \label{eq:normal-A01}
A_{01} &= \frac{1}{\phi(\mu; \mu_p, \sigma_p^2)} \times
\frac{\pi_1 \phi(\mu; 0,1)}{\pi_0}  \nonumber \\
&= \frac{\pi_1 \sigma_p}{\pi_0} \times
\exp\biggl\{ -\frac{\mu^2}{2} + \frac{(\mu - \mu_p)^2}{2\sigma_p^2} \biggr\} .
\end{align}
The reverse $1\to 0$ move is accepted with probability $\min\{1, A_{10}\}$, where $A_{10}
= \frac{1}{A_{01}}$.  The A1 algorithm includes a within-model step, which is simply a
round of sampling from the target $\mu \sim \N(0, 1)$ when the chain is in $H_1$.

\textit{\bf (i) The case of $\pi_0 \le \frac{1}{2}$} (e.g., $\pi_0=0.3$ in
fig.~\ref{fig:normal-pjump-eff}\textbf{a}--\textbf{c}).  The maximum $\Pj^* = 2\pi_0$ is
achieved by having $A_{01} \ge 1$ for all $\mu$, or if the following inequality holds
for all $\mu$:
\begin{equation} \label{eq:normal-A01-condition}
(1 - \sigma_p^2) \mu^2 - 2\mu_p\mu + \mu_p^2 + 2\sigma_p^2 \log \frac{\pi_1\sigma_p}{\pi_0} \ge 0.
\end{equation}
The solution is $\sigma_p \le 1$ and $\Delta = \mu_p^2 - 2(1 - \sigma_p^2) \log
\frac{\pi_1\sigma_p}{\pi_0} \le 0$, which are equivalent to the conditions
\begin{equation}\label{eq:N-round-table-pi<0.5}
\frac{\pi_0}{\pi_1} \le \sigma_p \le 1, \quad
\bigl| \mu_p \bigr| \le \sqrt{2(1 - \sigma_p^2) \log \frac{\pi_1\sigma_p}{\pi_0}}.
\end{equation}
This region is represented by the round flat top of the table mountain in
fig.~\ref{fig:normal-pjump-eff}\textbf{a} for the case of $\pi_0=0.3$.  Note that the
region represents a class of MCMC algorithms.  For each A0 algorithm, every $0\to 1$
proposal is accepted as $A_{01} \ge 1$ for all $\mu$, so that the proposal density and the
entrance distribution match, but both differ from the posterior, $g(\cdot) =
f_{01}^\text{in}(\cdot) \ne \pi_1(\cdot)$.  Even though maximum $\Pj$ is achieved by the
whole class of A0 algorithms, $E$ is not optimal except when $(\mu_p, \sigma_p) = (0,1)$.
Including the within-model move in the A1 algorithm to restore $\mu$ to its stationary
distribution achieves $E^*$ for the whole class of algorithms.

When $(\mu_p,\sigma_p)$ is outside the region of eq.~\ref{eq:N-round-table-pi<0.5},
$A_{01}>1$ for some values of $\mu$ and $<1$ for others.  The value of $\mu$ at which
$A_{01} = 1$ is given by the equation
\begin{equation} \label{eq:mu12N}
(1 - \sigma_p^2) \mu^2 - 2\mu_p\mu + \mu_p^2 + 2\sigma_p^2 \log \frac{\pi_1\sigma_p}{\pi_0} = 0.
\end{equation}
When $\sigma_p=1$ and $\mu_p\neq 0$, eq.~\ref{eq:mu12N} has only one solution, i.e.,
$\mu_1^* = \mu_2^* = \frac{\mu_p^2 + 2\log(\pi_1/\pi_0)}{2\mu_p}$. When $\sigma_p\neq 1$,
there are two solutions: $\mu_{1,2}^* = \frac{\mu_p\mp \sigma_p
	\sqrt{\Delta}}{1-\sigma_p^2}$.

Thus if $0<\sigma_p<\frac{\pi_0}{\pi_1}$ and $-\infty < \mu_p < \infty$, or
$\frac{\pi_0}{\pi_1} \le \sigma_p < 1$ and $|\mu_p| > \sqrt{2(1 - \sigma_p^2) \log
	\frac{\pi_1\sigma_p}{\pi_0}}$,
\begin{align} \label{eq:normal-Pj-a}
\Pj &= 2\pi_0 \Bigl[\Phi\bigl(\tfrac{\mu_1^*-\mu_p}{\sigma_p}\bigr)
+ \Phi\bigl(\tfrac{\mu_p-\mu_2^*}{\sigma_p}\bigr)
\Bigr] \nonumber \\
&+ 2\pi_1[\Phi(\mu_2^*) - \Phi(\mu_1^*)],
\end{align}
where $\Phi(\mu) = \int_{-\infty}^\mu \phi(\xi) \d \xi$ is the CDF for
$\N(0,1)$. The entrance distribution is
\begin{equation} \label{eq:normal-entrance-a}
f_{01}^\text{in}(\mu) =
\begin{cases}
\frac{\pi_1 \pi_1(\mu)/\pi_0}{z},  & \text{if } \mu_1^* \le \mu \le \mu_2^*, \\
\frac{g(\mu)}{z},                  & \text{otherwise},
\end{cases}
\end{equation}
where $z = \frac{1}{2\pi_0}\Pj = \Phi\bigl(\frac{\mu_1^* - \mu_p}{\sigma_p}\bigr) +
\Phi\bigl(\frac{\mu_p - \mu_2^*}{\sigma_p}\bigr) + \frac{\pi_1}{\pi_0} \bigl[\Phi(\mu_2^*) -
\Phi(\mu_1^*) \bigr]$.

If $\sigma_p=1$ and $\mu_p<0$,
\begin{equation} \label{eq:normal-Pj-b}
\Pj = 2\pi_0 \Phi\bigl( \tfrac{\mu_p-\mu_1^*}{\sigma_p} \bigr) + 2\pi_1 \Phi(\mu_1^*),
\end{equation}
and the entrance distribution is
\begin{equation} \label{eq:normal-entrance-b}
f_{01}^\text{in}(\mu) =
\begin{cases}
\frac{\pi_1 \pi_1(\mu)/\pi_0}{z},  & \text{if } \mu \le \mu_1^*, \\
\frac{g(\mu)}{z},                  & \text{otherwise},
\end{cases}
\end{equation}
where $z = \Pj/(2\pi_0)=\Phi\bigl(\frac{\mu_p-\mu_1^*}{\sigma_p} \bigr) +
\frac{\pi_1}{\pi_0}\Phi(\mu_1^*)$.

If $\sigma_p=1$ and $\mu_p>0$,
\begin{equation} \label{eq:normal-Pj-c}
\Pj = 2\pi_0 \Phi\bigl(\tfrac{\mu_1^* - \mu_p}{\sigma_p}\bigr) + 2\pi_1 \Phi(-\mu_1^*) ,
\end{equation}
and the entrance distribution is
\begin{equation} \label{eq:normal-entrance-c}
f_{01}^\text{in}(\mu) =
\begin{cases}
\frac{g(\mu)}{z},                  & \text{if } \mu < \mu_1^*, \\
\frac{\pi_1 \pi_1(\mu)/\pi_0}{z},  & \text{otherwise},
\end{cases}
\end{equation}
where $z = \Pj/(2\pi_0) = \Phi\bigl( \frac{\mu_1^* - \mu_p}{\sigma_p} \bigr) +
\frac{\pi_1}{\pi_0}\Phi(-\mu_1^*)$.

If $\sigma_p >1$ and $-\infty < \mu_p < \infty$, we have
\begin{align} \label{eq:normal-Pj-d}
\Pj &= 2\pi_0
\Bigl[\Phi\bigl(\tfrac{\mu_1^* - \mu_p}{\sigma_p}\bigr) -
\Phi\bigl(\tfrac{\mu_2^* - \mu_p}{\sigma_p}\bigr)
\Bigr]                         \nonumber \\
&+ 2\pi_1[\Phi(\mu_2^*) + \Phi(-\mu_1^*)] ,
\end{align}
and the entrance distribution is
\begin{equation} \label{eq:normal-entrance-d}
f_{01}^\text{in}(\mu) =
\begin{cases}
\frac{g(\mu)}{z},                   & \text{if } \mu_2^* \le \mu \le \mu_1^*, \\
\frac{\pi_1\pi_1(\mu)/\pi_0}{z},    & \text{otherwise},
\end{cases}
\end{equation}
where $z = \Phi\bigl(\frac{\mu_1^* - \mu_p}{\sigma_p}\bigr)
- \Phi\bigl(\frac{\mu_2^* - \mu_p}{\sigma_p}\bigr)
+ \frac{\pi_1}{\pi_0} [\Phi(\mu_2^*) + \Phi(-\mu_1^*) ]$.

In all those cases, where the proposal is outside the round top of the mountain
(eq.~\ref{eq:N-round-table-pi<0.5}), algorithm A1 including the within-model move has
higher efficiency $E$ than algorithm A0 without the within-model move, but neither
algorithm achieves the maximum $\Pj$ or maximum $E$.

\textit{\bf (ii) The case of $\pi_0 > \frac{1}{2}$} (e.g., $\pi_0 = 0.7$ in
fig.~\ref{fig:normal-pjump-eff}\textbf{g}--\textbf{i}).  The condition for achieving the
maximum $\Pj^* = 2\pi_1$ is
\begin{equation}\label{eq:N-round-table-pi>0.5}
1 \le \sigma_p \le \frac{\pi_0}{\pi_1},  \quad %
\bigl| \mu_p \bigr| \le \sqrt{2(1 - \sigma_p^2) \log \frac{\pi_1\sigma_p}{\pi_0}} .
\end{equation}
For the case of $\pi_0 = 0.7$, this region is represented by the round flat top of the
table mountain in figure \ref{fig:normal-pjump-eff}\textbf{g}\&\textbf{i}.  In this case,
$A_{01} \le 1$ for all $\mu$, the entrance distribution matches the posterior, with
$g(\cdot) \ne f_{01}^\text{in}(\cdot) = \pi_1(\cdot)$, and the A0 algorithm without
within-model move achieves maximum $\Pj$ and maximum $E$.  Including the within-model move
in A1 then does not improve efficiency any further.

When $(\mu_p, \sigma_p)$ is outside the region of eq.~\ref{eq:N-round-table-pi>0.5},
$A_{01}>1$ for some values of $\mu$ and $<1$ for others.  The value of $\mu$ at which
$A_{01} = 1$ is given by eq.~\ref{eq:mu12N}.  Thus if $0 < \sigma_p < 1$ and $-\infty <
\mu_p < \infty$, $\Pj$ is given by eq.~\ref{eq:normal-Pj-a} and the entrance distribution
by eq.~\ref{eq:normal-entrance-a}.  If $\sigma_p=1$ and $\mu_p<0$, $\Pj$ is given by
eq.~\ref{eq:normal-Pj-b} and the entrance distribution by eq.~\ref{eq:normal-entrance-b}.
If $\sigma_p=1$ and $\mu_p>0$, $\Pj$ is given by eq.~\ref{eq:normal-Pj-c} and the entrance
distribution by eq.~\ref{eq:normal-entrance-c}. If $1 < \sigma_p \le \frac{\pi_0}{\pi_1}$
and $|\mu_p| > \sqrt{2(1 - \sigma_p^2) \log \frac{\pi_1\sigma_p}{\pi_0}}$ or if
$\sigma_p>\frac{\pi_0}{\pi_1}$ and $-\infty < \mu_p < \infty$, $\Pj$ is given by
eq.~\ref{eq:normal-Pj-d} and the entrance distribution by eq.~\ref{eq:normal-entrance-d}.

With those proposals, outside the region of eq.~\ref{eq:N-round-table-pi>0.5}, algorithms
A0 and A1 do not achieve maximum $\Pj$ or maximum $E$, and including within-model moves in
A1 improves the mixing efficiency $E$ compared with algorithm A0.  See figure
\ref{fig:normal-pjump-eff}\textbf{g}--\textbf{i} for the case of $\pi_0=0.7$.

\newpage
\section{SI text 4. Super-efficient Markov chains in the discrete case} 

Here we provide an overview of MCMC mixing efficiency in the discrete case. 
\citet{Frigessi1992}, in their Proposition 1, characterized the most efficient
$\pi$-reversible Markov chain in the case where the function $f$ is an eigenvector of
$P$. We provide a simplified version of the proof of \citet{Frigessi1992}, using basic
linear algebra and avoiding the use of the inverse of the singular matrix, $(I -
P)^{-1}$.  The state space of the Markov chain is $\{ 1,2,\cdots, K \}$, which may
correspond to $K$ models with no parameters.  The transition matrix $P$ is irreducible
and reversible with target distribution $\pi = (\pi_1, \cdots, \pi_K )^T$, with $\pi_k >
0$ for every $k$.

\subsection{Spectral decomposition of $P$ and asymptotic variance}

Define $B = \mathrm{diag}(\pi_1, \cdots, \pi_K)$ and $\Pi = P^\infty = \mathbf{1} \pi^T$,
where $\mathbf{1}$ is the $K\times 1$ vector with all elements to be 1 and the superscript
`$T$' means transpose.  $\Pi$ is the limiting matrix of $P^n$ with $n\to \infty$ and
represents the independent sampler.  Note also $B\cdot\bf{1} = \pi$ and $\Pi\cdot\bf{1} =
\bf{1}$.  Because of the detailed-balance condition, $S = B^{\frac{1}{2}} P
B^{-\frac{1}{2}}$ is symmetrical and thus diagnosable. Let $S = R \Lambda R^T$, with
$\Lambda = \mathrm{diag}(\lambda_1, \lambda_2, \cdots, \lambda_K)$ to be the eigenvalues
of $S$ (which are all real), and columns of $R$ to be the corresponding right
eigenvectors, with $R^T = R^{-1}$.  Then
\begin{equation} \label{eq:P-spectral}
P = B^{-\frac{1}{2}} S B^{\frac{1}{2}}
= (B^{-\frac{1}{2}} R) \Lambda (R^T B^{\frac{1}{2}}) = E \Lambda E^T B,
\end{equation}
with $E = B^{-\frac{1}{2}} R$ and $E^T B E = I$.  Thus $\Lambda$ are eigenvalues of $P$,
while the columns of $E$ are the corresponding right eigenvectors.  As $P$ is an
irreducible stochastic matrix, we have $1 = \lambda_1 > \lambda_2 \ge \cdots \ge
\lambda_K > -1$.  Eq.~\ref{eq:P-spectral} is known as the spectral decomposition of $P$.
Any algebraic function of $P$, say $h(P)$, is then given as
\begin{equation}\label{eq:hP}
h(P) = E h(\Lambda) E^T B = E \,\mathrm{diag}\{h(\lambda_1), \cdots, h(\lambda_K)\} E^T
B.
\end{equation}

The asymptotic variance $\nu$ for estimating $\E_\pi(f) = \sum_k \pi_k f_k $ can then be
given in the matrix notation as \citep{Kemeny1960, Peskun1973, Green1992}
\begin{equation} \label{eq:nu-matrix}
\nu = \nu(f,\pi,P) = f^T B \bigl[ 2(I - P + \Pi)^{-1} - I - \Pi \bigr] f,
\end{equation}
where $Z=(I - P + \Pi)^{-1}$ is known as the \textit{fundamental matrix} for $P$
\citep{Kemeny1960, Peskun1973}.  The term in the square brackets is a function of $P$
(note that $I = P^0, \Pi = P^\infty$), and has eigenvalues $\delta_1 = 2(1 - \lambda_1 +
1)^{-1} - 1 - 1 = 0$ and $\delta_k = 2(1-\lambda_k)^{-1} - 1 =
\frac{1+\lambda_k}{1-\lambda_k}$ for $k \ge 2$. Thus eq.~\ref{eq:nu-matrix} can also be
written as \citep{Sokal1989, Frigessi1992, Green1992}
\begin{equation} \label{eq:nu-spectral}
\begin{aligned}
\nu &= f^T B E \cdot \text{diag}(\delta_1, ..., \delta_K) \cdot E^TBf \\
&= \sum_{k \ge 2} \frac{1+\lambda_k}{1-\lambda_k} (E^T B f)_k^2,
\end{aligned}
\end{equation}
where $(E^T B f)_k$ is the $k$-th element in the vector $E^T B f$. 
Eq.~\ref{eq:nu-spectral} highlights the importance of the eigenvalues, with small
eigenvalues representing high mixing efficiency.  In particular $1 - \lambda_2$ is known
as the \textit{spectral gap} or \textit{eigenvalue gap}, and a large spectral gap means
good mixing.

The question addressed here is the following: given $\pi$ and $f$, what is the
$\pi$-reversible stochastic matrix $P$ that minimizes $\nu$?  Note that minimizing $\nu$
(eq.~\ref{eq:nu-matrix}) is equivalent to minimizing
\begin{equation}
\varphi(P) = f^T B (I - P + \Pi)^{-1} f.
\end{equation}

\subsection{Directional derivatives}

We consider the change in $\nu$ or $\varphi$ when $P$ is perturbed by a small amount in
the direction of $H$ such that $P + \delta H$ remains an irreducible $\pi$-reversible
stochastic matrix.  Such a direction may be called a feasible direction.  We consider
feasible directions because we focus on Markov chains generated by the
Metropolis-Hastings algorithm, which are reversible with stationary distribution $\pi$. 
\citet{Frigessi1992} pointed out that the set of irreducible $\pi$-reversible stochastic
matrices is a convex set; in other words, any matrix between two matrices in the set is
in the set as well.  In particular, if both $P$ and $P + \delta H$ are in the set, $P +
\alpha\delta H$ for any $\alpha \in (0,1)$ will be in the set also.  The directional
derivative of $\varphi(P)$ at $P$ in the direction $H$ is
\begin{equation} \label{eq:varphi'}
\small
\begin{aligned}
\varphi'_H(P)
&= f^T B \cdot
\lim_{\delta \to 0} \frac{(I - P - \delta H + \Pi)^{-1} - (I - P + \Pi)^{-1}}{\delta}
\cdot f \\
&= f^T B \cdot [(I - P + \Pi)^{-1} H (I - P + \Pi)^{-1}] \cdot f.
\end{aligned}
\end{equation}
Here note that for any invertible matrix $X$,
\begin{equation}
\lim_{\delta \to 0} \frac{(X + \delta H)^{-1} - X^{-1}}{\delta} = -X^{-1} H X^{-1}.
\end{equation}
This can be shown as follows.  Let
\begin{equation}
\frac{(X + \delta H)^{-1} - X^{-1}}{\delta} = D.
\end{equation}
Then
\begin{equation}
(X + \delta H)^{-1} = X^{-1} + \delta D.
\end{equation}
Left-multiplying both sides by $(X + \delta H)$ gives
\begin{equation}
I = I + \delta H X^{-1} + \delta X D  + \delta^2 H D.
\end{equation}
Let $\delta \to 0$ and ignore the term involving $\delta^2$, and we get $D = -X^{-1} H
X^{-1}$.

Now suppose there exists an irreducible $\pi$-reversible $P$ which has $f$ as an
eigenvector, with $Pf = \lambda f$, $\bigl| \lambda \bigl|  < 1$.  Then
\begin{equation}
(I - P + \Pi)^{-1} \cdot f 
= (\lambda^0 - \lambda + \lambda^\infty)^{-1} f
= \tfrac{1}{1 - \lambda} \cdot f,   \nonumber
\end{equation}
and
\begin{equation}
f^T B (I - P + \Pi)^{-1} = [ B \cdot (I - P + \Pi)^{-1} f] ^T
= \tfrac{1}{1 - \lambda} \cdot f^T B,   \nonumber
\end{equation}
since $BP$ and $B (I - P + \Pi)^{-1}$ are symmetrical matrices, due to detailed balance.

Eq.~\ref{eq:varphi'} then becomes
\begin{equation} \label{eq:psi-phi}
\varphi'_H(P) = \frac{1}{(1 - \lambda )^2} f^T B H f
= \frac{1}{(1 - \lambda )^2} \psi'_H(P),
\end{equation}
where $\psi'_H(P) = f^T B H f$ is the directional derivative of $\psi(P) = f^T B P f$ at
$P$ in the direction $H$.

Thus $\varphi'_H(P) \ge 0$ if and only if $\psi'_H(P) \ge 0$.  A necessary and
sufficient condition for $P$ to attain a minimum of $\varphi(P)$ is
\begin{equation} \label{eq:varphi-condition}
\varphi'_H(P) \ge 0,  \ \  \text{or equivalently}, \ \  \psi'_H(P) \ge 0,
\end{equation}
for any feasible direction $H$.  If all feasible directions are upwards (with positive
slope), the current point ($P$) must be a minimum.  In the case where $P$ can have $f$ as
an eigenvector, minimization of $\nu$ is equivalent to minimization of $\psi(P)$ or
$\varphi(P)$.  Estimation of any model probability, say $\pi_1$, is such a case.

\subsection{Estimation of $\pi_1$ using $P$ constructed with orthogonal projection}

Eq.~\ref{eq:psi-phi} is derived under the assumption that there exists an irreducible
$\pi$-reversible $P$ which has $f$ as an eigenvector.  One important such case is the
estimation of the probability for one state, say, $\pi_1$.  The theory does not require
$\pi_1$ to be either the largest or the smallest.  The function may be $f =
(1,0,\cdots,0)^T$, but we `center' it to have zero mean,
\begin{equation} \label{eq:f}
f = ( 1-\pi_1, -\pi_1, \cdots, -\pi_1 )^T,
\end{equation}
with $\E_\pi(f) = \sum_k \pi_k f_k = 0$, and variance $\V_\pi(f) = \sum_k \pi_k f_k^2$ $=
\pi_1(1-\pi_1) \equiv \nu_f$.  We also write $\sum_k \pi_k f_k^2 \equiv \| f \| _\pi^2$.

As
\begin{equation}
Pf = \bigl[
p_{11}-\pi_1, \tfrac{\pi_1}{\pi_2}p_{12} - \pi_1, \cdots, \tfrac{\pi_1}{\pi_K}p_{1K} - \pi_1
\bigr]^T,
\end{equation}
we have
\begin{equation} \label{eq:psi}
\begin{aligned}
\psi(P) &= f^T BPf \\
&= \sum_k \pi_k \cdot f_k \cdot (Pf)_k  \\
&= (p_{11} - \pi_1)\pi_1 (1 - \pi_1)
- \pi_1 \sum_{k=2}^K \bigl( \tfrac{\pi_1}{\pi_k}p_{1k} - \pi_1 \bigr) \pi_k  \\
&=  (p_{11} - \pi_1)\pi_1.
\end{aligned}
\end{equation}
To find the minimum $p_{11}$, note that
\begin{equation} \label{eq:p11}
\begin{aligned}
p_{11} &= 1 - \sum_{j\ne 1} p_{1j}
= 1 - \sum_{j\ne 1} \tfrac{\pi_j}{\pi_1}p_{j1} \\
&\ge 1 - \sum_{j\ne 1} \tfrac{\pi_j}{\pi_1}
= \tfrac{2\pi_1-1}{\pi_1}.
\end{aligned}
\end{equation}

If $\pi_1 > \frac{1}{2}$ (in which case model 1 must be the MAP model), this bound is
reached by having $p_{j1} = 1$ (and $p_{1j} = \pi_j / \pi_1$) for all $j \ge 2$, giving
a unique matrix
\begin{equation}  \label{eq:P-pi>1/2-SI}
P^* =
\begin{bmatrix}
\frac{2\pi_1-1}{\pi_1} & \frac{\pi_2}{\pi_1} & \cdots & \frac{\pi_K}{\pi_1} \\
1                      & 0                     & \cdots & 0 \\
1                      & 0                     & \cdots & 0 \\
\vdots                 & \vdots                & \ddots & \vdots \\
1                      & 0                     & \cdots & 0
\end{bmatrix}.
\end{equation}
This is eq.~\ref{eq:P-pi>1/2}, which gives the maximum $P_{\mathrm{jump}} = 2(1-\pi_1)$
and maximum efficiency $E = 1/(2\pi_1 - 1)$. It is also easy to confirm that
\begin{equation}
P^* f = -\tfrac{1-\pi_1}{\pi_1} f,
\end{equation}
so that $P^*$ has $f$ as an eigenvector with eigenvalue $\lambda =
-\frac{1-\pi_1}{\pi_1}$.

In the case of $\pi_1 < \frac{1}{2}$, the bound based on $p_{11}$ (eq.~\ref{eq:p11}) is
ineffective.  The optimal $P$ can be constructed as follows.  Let $A = \{ a_{ij} \}$,
with $a_{ij} = \pi_j f_i f_j$, be the orthogonal projection operator in the direction of
$f$. The row sums of $A$ are all 0 and $\pi_i a_{ij} = \pi_j a_{ji}$ so that $A$ is
$\pi$-reversible.  Define $P_\epsilon = \Pi - \epsilon A$ with $\epsilon > 0$. Then
$P_\epsilon$ will be a $\pi$-reversible stochastic matrix as long as all its elements
are nonnegative, or if $\epsilon \le 1/\mathrm{max}_i f_i^2$.

Note that $P_{\epsilon}$ has $f$ as an eigenvector with eigenvalue $\lambda =$
$-\epsilon\| f \| _\pi^2$, because $(\Pi f)_j = 0$ and $(\epsilon A f)_j
= \epsilon \sum_k \pi_k f_j f_k^2 = \epsilon f_j \| f \|_\pi^2$, and
\begin{equation}
P_\epsilon f = \Pi f - \epsilon A f = -\epsilon Af = -\epsilon \| f \| _\pi^2 \cdot f.
\end{equation}

Let $P^*$ be $P_\epsilon$ with $\epsilon$ at its largest allowed value
\begin{equation}\label{eq:epsilon}
\epsilon_* = 1/\mathrm{max}_i \, f_i^2
=
\begin{cases}
1/\pi_1^2,      & \text{if } \pi_1 \ge 1/2, \\
1/(1-\pi_1)^2,  & \text{otherwise}.
\end{cases}
\end{equation}
This has the eigenvalue 
\begin{equation}\label{eq:lambda-epsilon}
\lambda_* = -\epsilon_* \| f \| _\pi^2
=
\begin{cases}
-\nu_f/\pi_1^2,      & \text{if } \pi_1 \ge 1/2, \\
-\nu_f/(1-\pi_1)^2,  & \text{otherwise},
\end{cases}
\end{equation}
corresponding to the eigenvector $f$, while the other eigenvalues are 1 and 0 (with
multiplicity $K-2$).  Furthermore, 
\begin{equation} 
\psi(P^*) = \lambda_* \|f\|_{\pi}^2,
\end{equation}
and the asymptotic variance (eq.~\ref{eq:nu-matrix}) becomes
\begin{equation}
\nu(f, \pi, P^*) = \frac{1 + \lambda_*}{1 - \lambda_*} \times \nu_f
= | 2\pi_1 - 1| \cdot \nu_f,
\end{equation}
giving the efficiency (for estimating $\pi_1$) as $E = \nu_f/\nu = 1/| 2\pi_1 - 1|$.
Note that in eq.~\ref{eq:nu-spectral}, $e^T B f = 0$ for each eigenvector $e \ne f$, and
only the term corresponding to $\lambda_*$ is nonzero.

If $\pi_1 \ge \frac{1}{2}$, $P_\epsilon$ at maximum $\epsilon$ is $P^*$ of
eq.~\ref{eq:P-pi>1/2-SI}, given by minimizing $p_{11}$ or $\psi(P)$.

If $\pi_1 < \frac{1}{2}$, $P_\epsilon$ at maximum $\epsilon$ gives the following matrix
\begin{equation} \label{eq:P-pi<1/2-SI}
P^* =
\begin{bmatrix}
0                       & \frac{\pi_2}{1-\pi_1}              & \cdots & \frac{\pi_K}{1-\pi_1} \\
\frac{\pi_1}{1-\pi_1}   & \frac{1-2\pi_1}{(1-\pi_1)^2}\pi_2  & \cdots & \frac{1-2\pi_1}{(1-\pi_1)^2}\pi_K \\
\vdots                  & \vdots                             & \ddots & \vdots \\
\frac{\pi_1}{1-\pi_1}   & \frac{1-2\pi_1}{(1-\pi_1)^2}\pi_2  & \cdots & \frac{1-2\pi_1}{(1-\pi_1)^2}\pi_K
\end{bmatrix} .
\end{equation}
This is eq.~\ref{eq:P-pi<1/2}, which gives 
\begin{equation} \label{eq:Pj-pi<1/2}
\small
\Pj = \pi_1 + \sum_{k\ge 2} \pi_k \Big[ 1 - \tfrac{1-2\pi_1}{(1-\pi_1)^2}\pi_k \Bigr]
= 1 - \tfrac{1-2\pi_1}{(1-\pi_1)^2} \sum_{k\ge 2} \pi_k^2.
\end{equation} 
Note that $\Pj < 1$ even though $\Pj = 1$ may be easily achievable when $\pi_k <
\frac{1}{2}$ for all $k$.  In this case maximum $\Pj$ is neither necessary nor sufficient
for achieving maximum $E$.  When the Markov chain is in 1, it moves away immediately, to
other states in proportion to their probabilities.  When not in 1, the optimal chain visit
the other states according to their posterior and it is worse to jump too much.

\textbf{Remark.} Eq.~\ref{eq:psi-phi} is valid only if $P$ can have $f$ as an eigenvector.
$P_\epsilon$ constructed using the projection has $f$ as an eigenvector, but $P_\epsilon
+ \delta H$ may not.  This means that we can use any irreducible $\pi$-reversible
stochastic matrix as $H$, and scan the whole set of irreducible $\pi$-reversible
stochastic matrices.  Then $P^*$ is the optimal matrix with minimum $\nu$ among all
irreducible $\pi$-reversible stochastic matrices.

\vspace{2em}
\section{Extended Methods}

\subsection{Two datasets}

We tested different algorithms described in the paper using two datasets.  The first
consists of the $\psi\eta$-globin pseudogenes from six primate species: human,
chimpanzee, gorilla, orangutan, Rhesus monkey, and spider monkey
(fig.~\ref{fig:trees}\textbf{a}). The data are from \citet{Miyamoto1987} and used by
\citet{Rannala1996}.  There are 6,166 sites in the alignment.  The second dataset
consists of a segment of the mitochondrial genome that encodes cytochrome oxidase II and
cytochrome b from 36 mammalian species, with 1812 sites in the sequence alignment
(fig.~\ref{fig:trees}\textbf{b}).  The dataset is from \citet{Yang2003} and analyzed as
dataset 3 (M1510) in \citet{Lakner2008}.  We kept two of the nine mouse lemur species and
removed alignment gaps, so that the alignment had 29 species and 1,799 sites. The
modified dataset is referred to as M1510YY03 (fig.~\ref{fig:trees}\textbf{b}).

We used the JC model \citep{Jukes1969} in the Bayesian analysis.  Our purpose is to assess
the mixing efficiency of different cross-tree algorithms in estimating the posterior
probability for the MAP tree.  We used \textsc{MrBayes} \citep{Ronquist2012} to generate
the `correct' posterior (the gold standard), by averaging over three long runs.  For both
datasets, each run used two chains of $10^7$ iterations, sampling every 50 iterations.

We wrote a C program to implement the NNI and SPR algorithms discussed in the paper.  For
each algorithm, we ran the program 10 times, with the averages used to calculate $\Pj$ and
$E$. For the $\psi\eta$ dataset, the chain length was $10^6$ iterations after a burn-in of
$2\times 10^4$ iterations.  The starting tree was chosen at random.  For the mt dataset,
the starting tree was chosen from the top five trees with the highest posterior
probabilities, and each run involved $5 \times 10^7$ iterations, after a burn-in of $10^5$
iterations.  All runs were conducted on a computer server with Intel Xeon Gold 6154
processors.  While running times for different algorithms are comparable, they should be
taken as a rough guide as our implementation is a proof of concept and not optimized to
take advantage of well-known savings in Bayesian phylogenetics. %

\subsection{The baseline algorithm (SPR)}

Our baseline algorithm has two steps in each MCMC iteration, an SPR move to update the
tree topology and a within-tree move to update branch lengths.  The SPR move prunes off a
subtree and re-grafts it to the rest of the tree (called the backbone)
(fig.~\ref{fig:nni-spr}\textbf{b}\&\textbf{b'}).  First a branch $f$ is chosen at random
to be the focal branch.  If $f$ is internal, one of its two ends is picked at random to be
the pruned subtree ($S$).  Otherwise, the tip node is the subtree.  Then branch $f$
(together with subtree $S$) is pruned off and re-grafted at a random location in a
randomly chosen branch (say, branch $r$) on the backbone; here branches $x$ and $y$ are
excluded as targets to ensure a change to the tree topology.  The move merges two branches
$x$ and $y$ into one branch ($r'$) and splits one branch $r$ into two ($x', y'$).  The
move incurs a Hastings ratio $t_r/(t_x+t_y)$, where $t_x$ is the length of branch $x$,
etc.\ \citep[][p.287-9]{Yang2014}.

For the within-tree move, we sample a branch at random and change it via a multiplier,
which is equivalent to a uniform sliding window applied on the logarithmic scale.  The
burn-in is used to adjust the window size automatically, to achieve an acceptance rate of
$\sim 35\%$ \cite[][eq.~9]{Yang2013}.

We then consider variations to the baseline algorithm.

\subsection{Within-tree moves to update branch lengths with the tree fixed}

We consider five variants to the within-tree move.

\textit{A1. One $(2n-3)$-D move to change all branch lengths.}  All branch lengths are
updated simultaneously with a single $(2n-3)$-D move, applying a multiplier in each
dimension.  The same window size is applied in all dimensions, tuned to achieve an
acceptance rate of $\sim24\%$ (table 2 in \citealp{Thawornwattana2018BA}).

\textit{A2-1. A sequence of 1-D moves to change the branch lengths.}  We update
the branch lengths one by one, using a multiplier.  The same window size is applied to
all branch lengths, tuned to achieve an average acceptance rate of $\sim$40\%.

\textit{A2-2. A sequence of 1-D moves with an update of the tree length (sum of all
	branch lengths).}  This is the same as A2-1 except that we include a multiplier move to
update the tree length, which changes all branch lengths by the same factor.  The window
size is tuned to achieve an acceptance rate of $\sim$40\% since this is a 1-D move in
the ($2n-3$)-D space.

\textit{A3-1. A sequence of 1-D Bactrian-Laplace moves to change branch lengths.}  This
is the same as A2-1, except that we replace the uniform sliding window by a
Bactrian-Laplace (BL) move \citep{Yang2013, Thawornwattana2018BA}, which suppresses
proposals around the current value.  The window size is tuned to achieve an acceptance
rate of $\sim$30\% \citep{Yang2013}.

\textit{A3-2. A sequence of 1-D moves and a tree-length update using the
	Bactrian-Laplace proposal.}  This is the same as A3-1 except that we include a proposal
to change the tree length.  The window size is tuned to achieve an acceptance rate of
$\sim$30\%.

\subsection{Cross-tree moves to change the tree topology}

We explored several variants to the baseline algorithm, with different strategies to
propose alternative tree topologies and branch lengths for the new tree.

\textit{B1 (NNI$_0$) NNI with direct transfer of branch lengths.}  NNI is a special case
of SPR and favors smaller changes to the tree than SPR.  Each internal branch in an
unrooted tree (say $f$) defines the relationships among four subtrees ($a,b,c,d$), and
there are three possible trees connecting them
(fig.~\ref{fig:nni-spr}\textbf{a}\&\textbf{a'}).  The NNI move changes the current tree
into one of the two alternatives.  We sample an internal branch at random as the focal
branch $f$.  We then choose a subtree from $a$ and $b$ and another from $c$ and $d$, and
swap them to generate the new tree.  We transfer the branch lengths in the current tree
onto the new tree without any modification (so that the internal branch length $t_f$ in
the current tree becomes the internal branch length in the new tree, say).

\textit{B2 (NNI$_1$)  NNI with a modification of the focal branch length.} This is the
same as B1 except that the focal internal branch length is modified using a multiplier,
while other branch lengths are unchanged.  The window size (on the logarithmic scale) is
set to $\epsilon = 0.005$.

\textit{B3 (NNI$_5$) NNI with all five branch lengths around the focal branch modified
	using multipliers.} The same window size $\epsilon = 0.005$ is used in the five
dimensions.  This is called the stochastic NNI (stNNI) in \citet{Lakner2008}, although
stNNI proposes the three possible trees around branch $f$ with probability $\frac{1}{3}$
each, so that the move may change the branch lengths without modifying the tree.

\textit{B4 (NNI$_\text{bw}$) NNI with branch weights.}  We convert internal branch
lengths into weights to preferentially propose changes to the tree around short internal
branches \citep{Yang2014}.  This is used in the \textsc{bpp} program
\citep{Rannala2017,Flouri2018,Flouri2020}.  Internal branches are sampled according to
probabilities
\begin{equation} \label{eq:weight-branch}
w_f \propto t_f^{-1/2},
\end{equation}
for a branch with length $t_f$.  The branches to be swapped are chosen at random as in
B1.  After the target tree is selected, we transfer the branches across trees
without modification.  As branch lengths and branch weights are unchanged during the
move, the Hastings ratio is 1.

\textit{B5 (NNI$_\text{bw\&pw}$, NNI with branch weights and parsimony weights).}  This is
the same as B4 (NNI$_\text{bw}$) using branch weights (eq.~\ref{eq:weight-branch}) but in
addition uses parsimony scores as weights for choosing between the two candidate trees
(fig.~\ref{fig:nni-spr}).  Let $S_k$ be the parsimony score of the current tree $k$, and
$S_{k'}$ the score for a candidate tree $k'$.  A simple scheme \citep{Yang2014, Zhang2020}
is to sample candidate trees according to probabilities
\begin{equation} \label{eq:weight-parsimony}
w_{k'} \propto \e^{(S_{k} - S_{k'})/a}.
\end{equation}
We tested several values for $a$: $0.1, 0.5, 1, 5, 10, 20$, and used $a=0.5$ as it gave
good performance.  The move incurs a Hastings ratio $q_{k'k}/q_{kk'}$ due to the use of
weights.

\textit{B6 (NNI$_\text{local}$ move of \citet{Larget1999}).}  This is a variant of the NNI
move and merges two branches in the source tree into one branch and splits the target
branch in the target tree into two branches, as in the SPR move
(fig.~\ref{fig:nni-spr}\textbf{b}\&\textbf{b}$'$).  The move may not change the tree
topology.

\textit{B7 (SPR$_{\text{bw}}$) SPR with branch weights.}  This uses the same branch
weights as B4 (NNI$_\text{bw}$, eq.~\ref{eq:weight-branch}) and the rest is the same as in
the baseline algorithm.  The changes to branch lengths (due to splitting and merging
branches, fig.~\ref{fig:nni-spr}\textbf{b}\&\textbf{b}$'$) and the use of branch weights
incur a Hastings ratio.

\textit{B8 (SPR$_\text{bw\&pw}$) SPR with branch weights and parsimony weights.}  This is
the same as the baseline algorithm (B0) but uses branch weights
(eq.~\ref{eq:weight-branch}) to sample internal (focal) branches and parsimony weights
(eq.~\ref{eq:weight-parsimony}) for choosing among candidate branches for re-attachment.

\subsection{Division of computational efforts between within-tree and cross-tree moves}

Instead of having a within-tree move and a cross-tree move in each MCMC iteration, here
we sample the cross-tree move with probability $p_1$ (and the within-tree move with
$p_0 = 1-p_1$).

\FloatBarrier

\begin{figure} [t]
	\centering %
	\includegraphics[scale=0.58]{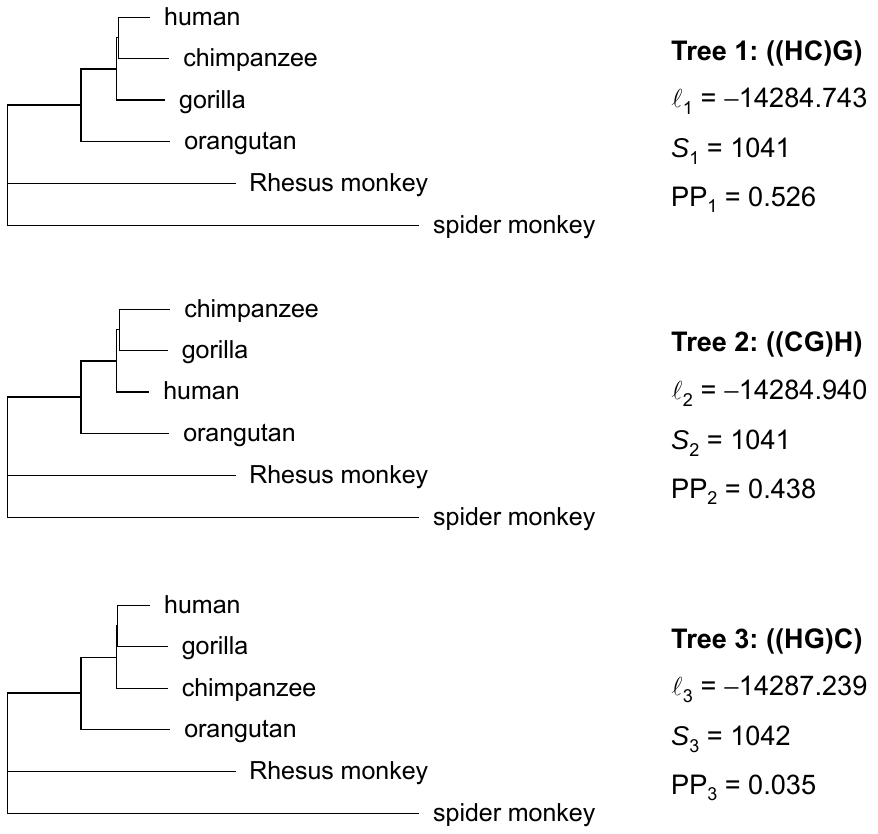} %
	
	\caption{Log likelihood values ($\ell$), parsimony scores (minimum number of changes,
		$S$), and Bayesian posterior probabilities (PP) for the top three trees for the
		$\psi\eta$ dataset.  The log likelihood values are in the same order as the posterior
		probabilities, but the top two trees have the same parsimony score.  The other trees
		have much worse scores using either of the three criteria.  The branches are drawn
		using the maximum likelihood estimates of their lengths. \\ %
	} \label{fig:trees-psieta-ml-mp}
\end{figure}

\begin{figure} [t]
	\centering %
	\includegraphics[scale=0.58]{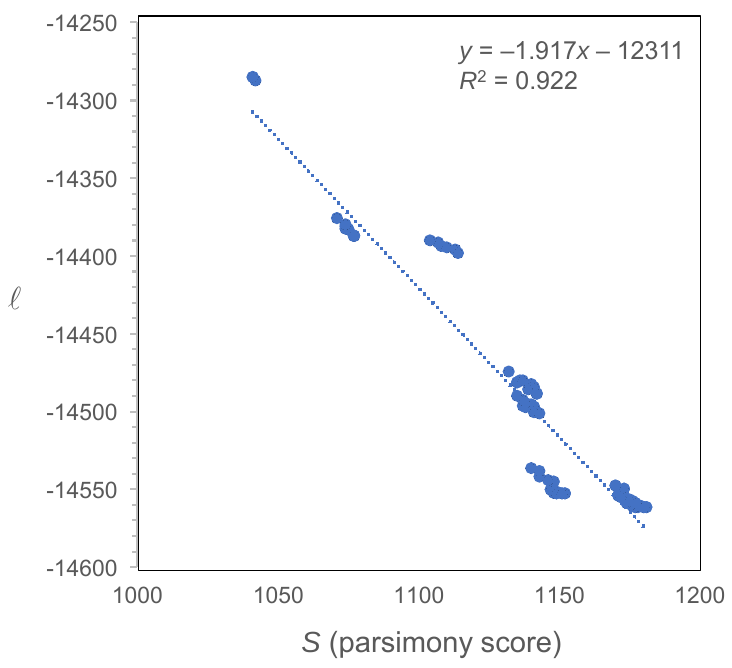} %
	
	\caption{Log-likelihood value ($\ell$) plotted against the parsimony score ($S$) for the
		105 trees for the $\psi\eta$ dataset.  The top three trees are shown in figure
		\ref{fig:trees-psieta-ml-mp}. \\ %
	} \label{fig:psieta-lnL-S}
\end{figure}

\end{document}